\documentclass[aps,prc,twocolumn,superscriptaddress,pdftex,nofootinbib]{revtex4-1}

\usepackage{amsmath}
\usepackage{amssymb}
\usepackage{listings}
\usepackage{dsfont}
\usepackage{slashed}
\usepackage{color}

\usepackage{graphicx}
\usepackage{epstopdf}
\usepackage{subfigure}
\usepackage{epsfig}
\usepackage{latexsym}
\usepackage{multirow}
\usepackage{mciteplus}
\usepackage{bm}

\usepackage{hhline}
\usepackage{lineno}

\def\vmed#1{\left< #1 \right>}

\usepackage{ulem}

\begin{document}

\title{Evaluation of the E2/M1 ratio in the $N\to  \Delta(1232)$
  transition from the $ \vec{\gamma} \vec{p} \to p \pi^0 $ reaction}

\author{E.~Mornacchi} \affiliation{Institut f\"ur Kernphysik, University of Mainz, D-55099 Mainz, Germany}
\author{P.~Pedroni} \affiliation{INFN Sezione di Pavia, I-27100 Pavia, Italy}
\author{F.~Afzal}
\author{Y.~Wunderlich} \affiliation{Helmholtz-Institut f\"ur Strahlen- und Kernphysik, Universit\"at Bonn, D-53115 Bonn, Germany}
\author{S.~Abt}\affiliation{Department f\"ur Physik, Universit\"at Basel, CH-4056 Basel, Switzerland}
\author{P.~Achenbach}\affiliation{Institut f\"ur Kernphysik, University of Mainz, D-55099 Mainz, Germany}
\author{J.R.M.~Annand}\affiliation{SUPA School of Physics and Astronomy, University of Glasgow, Glasgow G12 8QQ, United Kingdom}
\author{H.J.~Arends}\affiliation{Institut f\"ur Kernphysik, University of Mainz, D-55099 Mainz, Germany}
\author{M.~Bashkanov}\affiliation{Department of Physics, University of York, Heslington, York, Y010 5DD, UK}
\author{M.~Biroth}\affiliation{Institut f\"ur Kernphysik, University of Mainz, D-55099 Mainz, Germany}
\author{R.~Beck}\affiliation{Helmholtz-Institut f\"ur Strahlen- und Kernphysik, Universit\"at Bonn, D-53115 Bonn, Germany}
\author{N.~Borisov}\affiliation{Joint Institute for Nuclear Research, 141980 Dubna, Russia}
\author{A.~Braghieri}\affiliation{INFN Sezione di Pavia, I-27100 Pavia, Italy}
\author{W.J.~Briscoe}\affiliation{The George Washington University, Washington, DC 20052-0001, USA}
\author{F.~Cividini}\affiliation{Institut f\"ur Kernphysik, University of Mainz, D-55099 Mainz, Germany}
\author{C.~Collicott}\affiliation{Institut f\"ur Kernphysik, University of Mainz, D-55099 Mainz, Germany}
\author{A.~Denig}\affiliation{Institut f\"ur Kernphysik, University of Mainz, D-55099 Mainz, Germany}
\author{A.~S.~Dolzhikov}\affiliation{Joint Institute for Nuclear Research, 141980 Dubna, Russia}
\author{E.~Downie}\affiliation{The George Washington University, Washington, DC 20052-0001, USA}
\author{S.~Fegan}\affiliation{Department of Physics, University of York, Heslington, York, Y010 5DD, UK}
\author{A.~Fix}\affiliation{Tomsk Polytechnic University, 634034 Tomsk, Russia}
\author{D.~Ghosal}\affiliation{Department f\"ur Physik, Universit\"at Basel, CH-4056 Basel, Switzerland}
\author{I.~Gorodnov}\affiliation{Joint Institute for Nuclear Research, 141980 Dubna, Russia}
\author{W.~Gradl}\affiliation{Institut f\"ur Kernphysik, University of Mainz, D-55099 Mainz, Germany}
\author{G.~G.~Gurevich}\affiliation{Institute for Nuclear Research, 125047 Moscow, Russia}
\author{L.~Heijkenskj\"old}\affiliation{Institut f\"ur Kernphysik, University of Mainz, D-55099 Mainz, Germany}
\author{D.~Hornidge}\affiliation{Mount Allison University, Sackville, New Brunswick E4L 1E6, Canada}
\author{G.M.~Huber}\affiliation{University of Regina, Regina, Saskatchewan S4S 0A2, Canada}
\author{V.L.~Kashevarov}\affiliation{Institut f\"ur Kernphysik, University of Mainz, D-55099 Mainz, Germany}
\author{S.J.D.~Kay}\affiliation{Department of Physics, University of York, Heslington, York, Y010 5DD, UK}
\author{M.~Korolija}\affiliation{Rudjer Boskovic Institute, HR-10000 Zagreb, Croatia}
\author{A.~Lazarev}\affiliation{Joint Institute for Nuclear Research, 141980 Dubna, Russia}
\author{K.~Livingston}\affiliation{SUPA School of Physics and Astronomy, University of Glasgow, Glasgow G12 8QQ, United Kingdom}
\author{S.~Lutterer}\affiliation{Department f\"ur Physik, Universit\"at Basel, CH-4056 Basel, Switzerland}
\author{I.J.D.~MacGregor}\affiliation{SUPA School of Physics and Astronomy, University of Glasgow, Glasgow G12 8QQ, United Kingdom}
\author{D.M.~Manley}\affiliation{Kent State University, Kent, Ohio 44242-0001, USA}
\author{P.P.~Martel}\affiliation{Institut f\"ur Kernphysik, University of Mainz, D-55099 Mainz, Germany}
\author{R.~Miskimen}\affiliation{University of Massachusetts, Amherst, Massachusetts 01003, USA}
\author{M.~Mocanu}\affiliation{Department of Physics, University of York, Heslington, York, Y010 5DD, UK}
\author{C.~Mullen}\affiliation{SUPA School of Physics and Astronomy, University of Glasgow, Glasgow G12 8QQ, United Kingdom}
\author{A.~Neganov}\affiliation{Joint Institute for Nuclear Research, 141980 Dubna, Russia}
\author{A.~Neiser}\affiliation{Institut f\"ur Kernphysik, University of Mainz, D-55099 Mainz, Germany}
\author{M.~Oberle}\affiliation{Department f\"ur Physik, Universit\"at Basel, CH-4056 Basel, Switzerland}
\author{M.~Ostrick}\affiliation{Institut f\"ur Kernphysik, University of Mainz, D-55099 Mainz, Germany}
\author{P.B.~Otte}\affiliation{Institut f\"ur Kernphysik, University of Mainz, D-55099 Mainz, Germany}
\author{D.~Paudyal}\affiliation{University of Regina, Regina, Saskatchewan S4S 0A2, Canada}
\author{A.~Powell}\affiliation{SUPA School of Physics and Astronomy, University of Glasgow, Glasgow G12 8QQ, United Kingdom}
\author{T.~Rostomyan}\affiliation{Department f\"ur Physik, Universit\"at Basel, CH-4056 Basel, Switzerland}
\author{V.~Sokhoyan}\affiliation{Institut f\"ur Kernphysik, University of Mainz, D-55099 Mainz, Germany}
\author{K.~Spieker}\affiliation{Helmholtz-Institut f\"ur Strahlen- und Kernphysik, Universit\"at Bonn, D-53115 Bonn, Germany}
\author{O.~Steffen}\affiliation{Institut f\"ur Kernphysik, University of Mainz, D-55099 Mainz, Germany}
\author{I.I.~Strakovsky}\affiliation{The George Washington University, Washington, DC 20052-0001, USA}
\author{M.~Thiel}\affiliation{Institut f\"ur Kernphysik, University of Mainz, D-55099 Mainz, Germany}
\author{A.~Thomas}\affiliation{Institut f\"ur Kernphysik, University of Mainz, D-55099 Mainz, Germany}
\author{Yu.A.~Usov}\affiliation{Joint Institute for Nuclear Research, 141980 Dubna, Russia}
\author{S.~Wagner}\affiliation{Institut f\"ur Kernphysik, University of Mainz, D-55099 Mainz, Germany}
\author{D.P.~Watts}\affiliation{Department of Physics, University of York, Heslington, York, Y010 5DD, UK}
\author{D.~Werthm\"uller}\affiliation{Department of Physics, University of York, Heslington, York, Y010 5DD, UK}
\author{J.~Wettig}\affiliation{Institut f\"ur Kernphysik, University of Mainz, D-55099 Mainz, Germany}
\author{L.~Witthauer}\affiliation{Department f\"ur Physik, Universit\"at Basel, CH-4056 Basel, Switzerland}
\author{M.~Wolfes}\affiliation{Institut f\"ur Kernphysik, University of Mainz, D-55099 Mainz, Germany}
\author{N.~Zachariou}\affiliation{Department of Physics, University of York, Heslington, York, Y010 5DD, UK}

\noaffiliation
\collaboration{A2 Collaboration at MAMI}
\date{\today}

\begin{abstract}

A new  data set for the helicity-dependent differential cross section of the single-meson photoproduction reaction $\gamma p \to p \pi^{0}$ was obtained for the photon energy interval 150-400 MeV.
The experiment was performed at the A2 tagged photon facility of the Mainz Microtron MAMI 
using a circularly polarized photon beam and a longitudinally polarized proton target.
The reaction products were detected with the large acceptance Crystal Ball/TAPS calorimeter covering 97\% of the full solid angle.
These new results, obtained with a fine energy and polar angle binning, greatly increase both the existing quantity and quality of the data available for this observable.
A moment analysis, based on a finite expansion in Legendre polynomials, was applied to these data by using a
bootstrap-based fitting method to correctly account for their systematic uncertainties.
From the resulting decomposition of the differential cross sections,  
the $E2/M1$  ratio for the $N\to  \Delta(1232)$ transition was determined to be $[-2.38 \pm 0.16{\hbox{ (stat.+sys.)}} \pm 0.10 {\hbox{ (model)}}] \%$.
Combining this value with previous results also allowed us to evaluate the most precise available estimate of the $E2/M1$ ratio to be used for all further reference and model comparisons.
  

\end{abstract}

\maketitle

\section{Introduction}

Fundamental states of hadrons are characterized by
complex quark-gluon and meson cloud dynamics,
which are governed by non-perturbative QCD
and give rise to non-spherical components in their hadronic wave-function.
Precise experimental determination of the shapes then gives deep
insight into these interactions and provides a fundamental precision
benchmark for all types of models describing the hadronic structure
(see, for instance, Refs.~\cite{bern,vdh1} and references therein). 

The most direct and reliable evidence of a shape deformation is provided
by the determination of the particles' electric quadrupole moment.
However, the proton, the only stable hadron,
has a vanishing spectroscopic quadrupole moment in the
laboratory frame due to its spin-1/2 nature.
%
Therefore, one has to study the transition to its lowest $J=3/2$ excited state, 
namely the $\Delta(1232)$ resonance.   

The $N \to \Delta(1232)$ electromagnetic transition 
is  predominantly  due to  the  magnetic  dipole component $M1$.
In a very simple constituent quark model framework,
this process is described by a spin flip of a single quark in the $s$-wave state
(see, for instance, Ref.~\cite{Kru03} and references therein).
Using real photons, any $d$-wave mixture in the nucleon
and/or in the $\Delta(1232)$ wave functions allows for the electric quadrupole transition $E2$. 
Therefore,  by  measuring  the $\gamma p \to \Delta(1232) \to N\pi$ reactions,  
one can assess the presence of the $d$-wave components
and thus quantify to what extent the nucleon
and/or the $\Delta(1232)$ resonance deviate from the spherical shape.

The amplitudes in the $\gamma N\to N\pi$ final states are usually described by
the notation $E_{\ell\pm}^{I}$ and $M_{\ell\pm}^{I}$, where $E$ and $M$ are the electric
and magnetic multipoles, respectively, $\ell$ is the orbital angular momentum of the
photoproduced pion, the sign $\pm$ refers to the total $N\pi$
angular momentum $J=\ell\pm 1/2$,  and $I$ is the isospin of the
$N\pi$ system.

A common practice (see again Ref.~\cite{Kru03}) is to measure the resonant quadrupole strength
relative to the resonant dipole amplitude via the ratio

 \begin{equation}\label{eq:1new}
   R_{EM} = \frac{E2}{M1} \equiv
   \left.
   \frac{\hbox{Im}[E_{1+}^{3/2}]}{\hbox{Im}[M_{1+}^{3/2}]} \right\rvert_{M_\Delta}
 \end{equation}
evaluated at the $\Delta(1232)$ mass value $M_\Delta$.

%
 
Early empirical quark and Skyrme models, 
as well as relativistic models including two-body exchange currents,
gave  a variety of $R_{EM}$ values in the range $-6\% < R_{EM} < 0\%$, where
the minus sign indicates an  oblate shape deformation 
(see e.g., Refs.~\cite{beck1,sand,drec03} and references therein).

More recently, a similar range of values was predicted
by EFT approaches~\cite{vdh2,hemmert}, dynamical
and effective Lagrangian models~\cite{drec03,satolee,tjon,udias}, and
the most recent lattice calculations~\cite{alex3}. 


From an experimental point of view, 
isolating the resonant $E_{1+}$ amplitude is complicated by 
its rather small value compared to the dominant  $M_{1+}$ multipole transition.
%
To overcome these difficulties, 
high-intensity photon beams of precisely known energy are required,
along with the measurement of selected polarization observables
that highlight the role of the  small  $E_{1+}$ multipole thanks to the presence of  
interference terms between $E_{1+}$ and the dominant $M_{1+}$ multipole.
 
These conditions were met by the advent of a newer generation
of tagged photon facilities in the late 1990s.
In the last 20 years, several precision measurements have been performed
at the photon tagging facility of the MAMI Microtron in Mainz (Germany)~\cite{ahr04,beck2,beck3,galler} 
and at the laser back-scattering facility (LEGS) of the Brookhaven National Laboratory (USA)~\cite{sand}, 
where $R_{EM}$ was evaluated from both pion production and Compton scattering data.

The experimental apparatus of the LEGS Collaboration~\cite{sand}  had a complex geometry with a small angular acceptance and was best suited for photon detection. Measurements at the MAMI tagging photon facility have instead been performed over the years with different different detector setups: the large acceptance DAPHNE detector~\cite{beck2,ahr04}, covering the polar angle range $\theta_{\text{lab}} \in [21^{\circ},159^{\circ}]$ with full azimuthal acceptance and good charged particle detection capabilites, the TAPS calorimeter~\cite{beck2} and the LARA apparatus~\cite{galler},  both suited for the photon detection, which had a complex geometry and somewhat limited geometrical acceptance, although they covered more extreme forward and backward angles compared to the LEGS apparatus.

All of these experiments gave estimates of  $R_{EM}$ that are
compatible, within their errors, with the interval given
by the latest Particle Data Group (PDG) estimate~\cite{PDG22} of $ -3\% \le R_{EM} \le -2\%$.
A very similar interval ($R_{EM} =  -2.5\% \pm 0.4\% $)
has also been obtained from an analysis of the more recent $\gamma N \to N \pi$ data 
using the AMIAS methodology~\cite{amias}, a general-purpose algorithm applied for the  analysis of several different hadronic and  nuclear physics data sets~\cite{amias2,amias3}.

In this paper, we present a new precise determination of the $R_{EM}$ ratio from 
the measurement of the helicity-dependent differential cross section of the 
$\gamma p \to  p\pi^0$ reaction in the incident photon energy range from 150 to 400~MeV.
This experiment was performed at the A2 tagged photon facility of the MAMI electron accelerator
in Mainz, Germany~\cite{mamic}, using the 
experimental setup of the A2 Collaboration
that combines all the strengths of the previous experiments described above: a very large ($\sim 97\%$ of $4\pi$) angular 
acceptance with a good energy and angular resolution for photons and protons, as well as a high photon detection efficiency.
The measured double-polarization observable, accessed using a circularly polarized photon beam and
a longitudinally polarized proton target, can be defined as:

\begin{equation}
\frac{d\Delta \sigma}{d\Omega}= \Bigg(\frac{d\sigma}{d\Omega}\Bigg)_{3/2} - \Bigg(\frac{d\sigma}{d\Omega}\Bigg)_{1/2} \ , 
\label{E_obs1}
\end{equation}
where the subscripts $3/2$ and $1/2$ indicate the 
total helicity states of the $\gamma p$ system 
corresponding to the relative parallel or antiparallel photon-proton spin configurations, respectively.

Thanks to the very high statistics collected, it
was possible to measure
$d\Delta \sigma/d\Omega$  
in very fine beam photon energy 
and polar angle binning, with a width of $\simeq 2$~MeV and
$10^\circ$, respectively. 
%
This precise mapping of the entire $\Delta(1232)$ resonance excitation region
greatly enhanced both the existing quantity and quality
of the data available for this observable and 
allowed the existing estimates
of the $E2/M1$ ratio to be improved using a Legendre-moment  analysis.

The paper is organized as follows.
In Secs.~\ref{expapp} and~\ref{sec:ana},
the experimental setup and the offline analysis methods used to
obtain the measured observable are briefly described.
In Sec.~\ref{sec:results}, the new results
for $d\Delta \sigma/d\Omega$ are presented and compared with the scarce
existing database.
Section~\ref{sec:legendre_analysis} reports the Legendre-moment analysis of the
$d\Delta \sigma/d\Omega$ data, with the novel bootstrap-based fitting
method used in the current analysis described in Subsection~\ref{sec:fit}.
The determination of the $E2/M1$ ratio from 
the definitions of the fitted Legendre-moments in terms of multipoles
is reported in  Sec.~\ref{sec:e2m1}.
Finally, the summary and outlook are given in Sec.~\ref{sec:sum}.

\section{Experimental setup}\label{expapp}

The helicity-dependent data used for this analysis were collected at
the A2 tagged photon facility of the MAMI electron accelerator in Mainz, Germany~\cite{mamic}.

\begin{figure*}%
\centering
\includegraphics[scale=0.27]{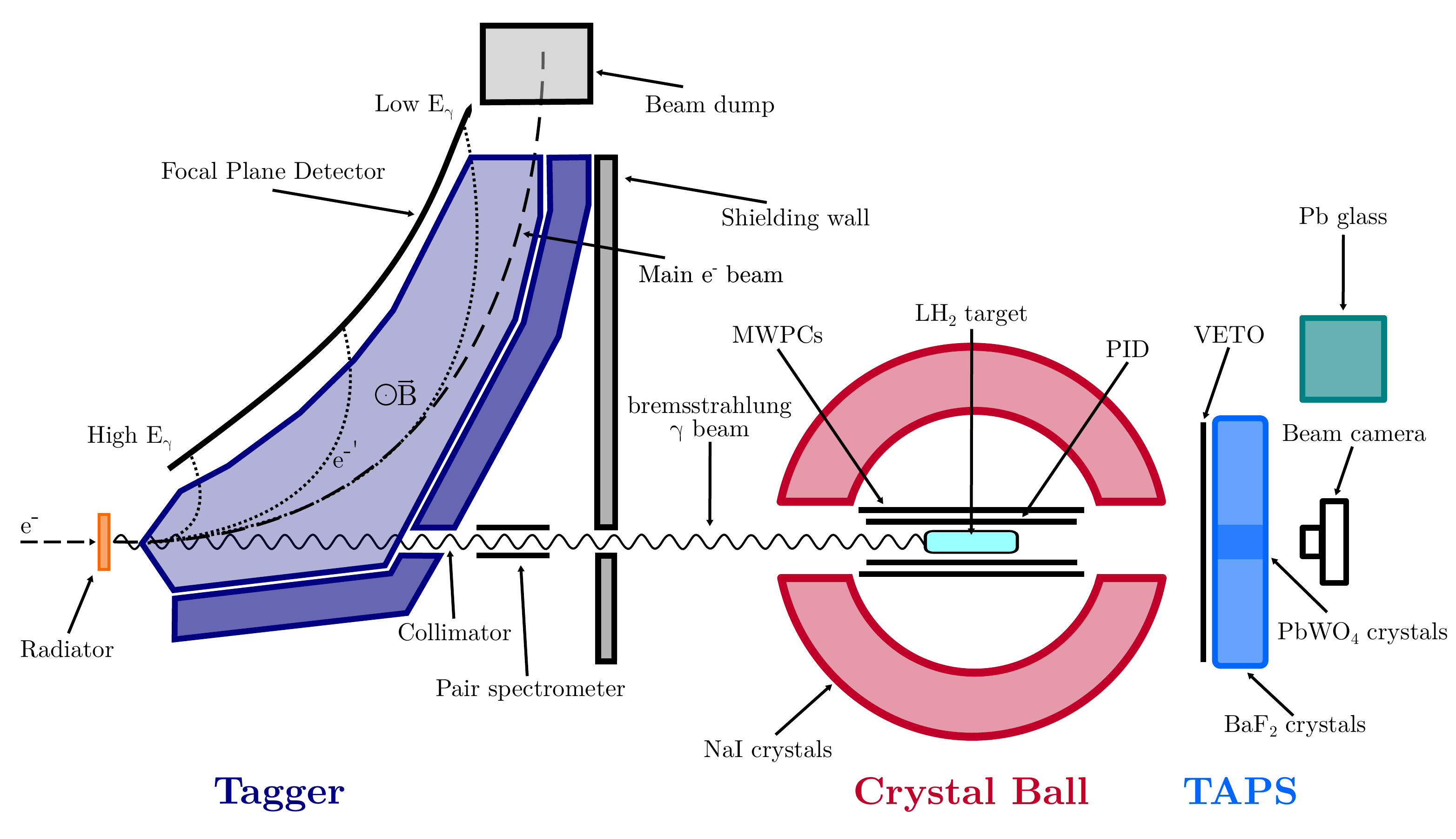}
\caption{Sketch of the experimental setup of the A2 Collaboration tagged photon facility, including photon tagging apparatus and detectors~\cite{edo}. The figure is not to scale.}
\label{fig_A2_setup}
\end{figure*}

Figure~\ref{fig_A2_setup} shows a sketch of the A2 experimental setup used for the measurement.
Since this setup has already been described in detail (see, for instance,
Refs.~\cite{A2:2013wkp,A2:2014pie,diet1,witt1,diet2,cividini} and
references therein), we will limit the discussion to the features relevant to the present experiment.


\subsection{The photon beam}

The circularly polarized photons used for this measurement were produced via Bremsstrahlung
on an amorphous radiator of the 450-MeV longitudinally polarized electron beam.
To avoid polarization dependent photon flux values,
the helicity of the electron beam was flipped at a rate of 1~Hz.

The electron polarization degree, $P_e$, was regularly
determined by Mott scattering close to the electron source~\cite{mott} 
and was found to be more than 80\% with a systematic uncertainty of $\pm 3\%$.

%
The recoil electrons from the Bremsstrahlung process were momentum-analyzed
using the Glasgow-Mainz spectrometer with an energy resolution of $\sim 1$~MeV,
corresponding to the width of the detector channels~\cite{taggnew}. 
The resulting photon beam passed through a 2~mm-diameter lead collimator,
reaching the target and detection apparatus.

The degree of the energy-dependent circular photon polarization, $P_{\odot}^{\gamma}$,
was determined using the Olsen and Maximon equation~\cite{olsen}:
\begin{equation*}
  \frac{P_{\odot}^{\gamma}}{P_e} = \frac{4x-x^2}{4-4x+3x^2} \ ,   
\end{equation*}
where $x=  E_e / E_\gamma$, with $E_e$ and $E_\gamma$ being the energy of the electron and the
Bremsstrahlung photon, respectively.     
 
The photon tagging efficiency was measured once a day using a Pb-Glass Cherenkov detector
in dedicated low flux runs.
During the standard data taking operation, fluctuations in the photon flux 
were monitored using a low-efficiency pair spectrometer 
located in the photon beamline after the collimator.
An absolute systematic uncertainty in the photon flux of 4\% was
estimated by comparing the data from these detectors obtained under
a range of different experimental conditions.

\subsection{The target system}

The longitudinally polarized proton target used in this experiment
was the Mainz-Dubna Frozen Spin Target~\cite{Rohl,Thomas}.
The filling factor for the $\sim 2$-mm-diameter butanol
spheres contained in the 2-cm-long, 2-cm-diameter target container
was estimated to be 60\%, with a systematic uncertainty of 2\%~\cite{Rohl}.

The target material (butanol) was polarized using the
Dynamic Nuclear Polarization effect~\cite{Brad99},
which requires a high magnetic field (about 2.5~T) and a temperature of about 25~mK.
A small holding magnetic field of 0.6~T, which replaced the polarizing magnet
during the data taking phase, allowed
regular relaxation times of about 1000~h to be achieved.

The target polarization was measured with an NMR system
before and after each data taking period and then
exponentially interpolated at intermediate times.
Corrections to the calculated polarization values were necessary due
to ice formation on the NMR coils.
They were taken from Refs.~\cite{dilli,A2:2019bqm}, which 
independently analyzed the same data set used in the present work.
Due to these corrections, as in Ref.~\cite{A2:2019bqm}, a conservative
systematic uncertainty of 10\% was applied to the target polarization values.

 
\subsection{The hadron detector}

The 
photon-induced reaction products 
were detected by
the Crystal Ball-TAPS apparatus.
The Crystal Ball (CB) calorimeter was placed around the target cell
and covered the full azimuthal ($\phi$) angle and a polar ($\theta$) angle range from 21$^{\circ}$ to 159$^{\circ}$~\cite{artcb}.
It consisted of 672 NaI(Tl) crystals and had a $\sim 100\%$ detection efficiency
for photons coming from the $\pi^0$ decay.
Inside CB, from the inside out, there were a Particle Identification Detector (PID),
consisting of a barrel of 24 plastic scintillators, and
two Multi-Wire Proportional Chambers (MWPCs).
The combination of all these detectors provided a precise tracking
and identification of charged particles. 
TAPS was an hexagonal wall covering the polar angle forward region outside the CB acceptance,
$1^{\circ} < \theta < 20^{\circ}$, and was made of of 366 BaF$_2$
and 72 PbWO$_{4}$ crystals~\cite{taps1,taps2}.
In front of each crystal there was a 5-mm-thick plastic scintillator (VETO)
that was used for charged particle identification.
The combination of the large acceptance CB and TAPS calorimeters covered $\sim$ 97$\%$ of the full solid angle. 
%
\section{Data Analysis} \label{sec:ana}

After offline energy and time calibration of all detectors,
the data from the butanol target were analyzed,
and all the implemented algorithms were tested and checked
with simulation, to obtain an optimal identification of the $\gamma p \to p \pi^0$ channel.

No unpolarized background had to be evaluated since the effect of the
$C$ and $O$ spinless nuclei present in the target vanishes in the $d\Delta\sigma/d \Omega$
difference (see Eq.~(\ref{E_obs1})).
%

\subsection{$\pi^{0}$ reconstruction and identification}

The algorithms used for the data selection were basically the same as those described in Ref.~\cite{cividini}.
Therefore, only a summary of the main analysis steps required to identify the $p \pi^0$ channel is given here.

The candidate events accepted for the differential cross section evaluation were those with 2 neutral or 3 (with at least two neutral) clusters reconstructed inside the detector. A neutral cluster is defined as an energy deposition in one of the two calorimeters, without an associated hit in either the PID or MWPC, or VETO for clusters in CB and TAPS, respectively.

Due to the relatively high tagged photon flux ($\approx~\!10^7\ \gamma/s$), a time coincidence within $20$~ns was required between 
the trigger in the calorimeters and the hits in the tagger focal plane detector. 
To remove the random coincidences in the selected time window, a side-band subtraction was also 
performed by selecting a background sample on each side of the prompt peak.

The first offline analysis step was the evaluation of the two-photon invariant mass (IM) using
all neutral clusters of each event. For all events with more than 2 neutral hits, all possible
combinations were used to calculate the IM and only the combination that gave
the closest value to the nominal $\pi^0$ mass was retained for subsequent analysis steps.

\begin{figure*}%
  \centering
  \includegraphics[scale=0.75]{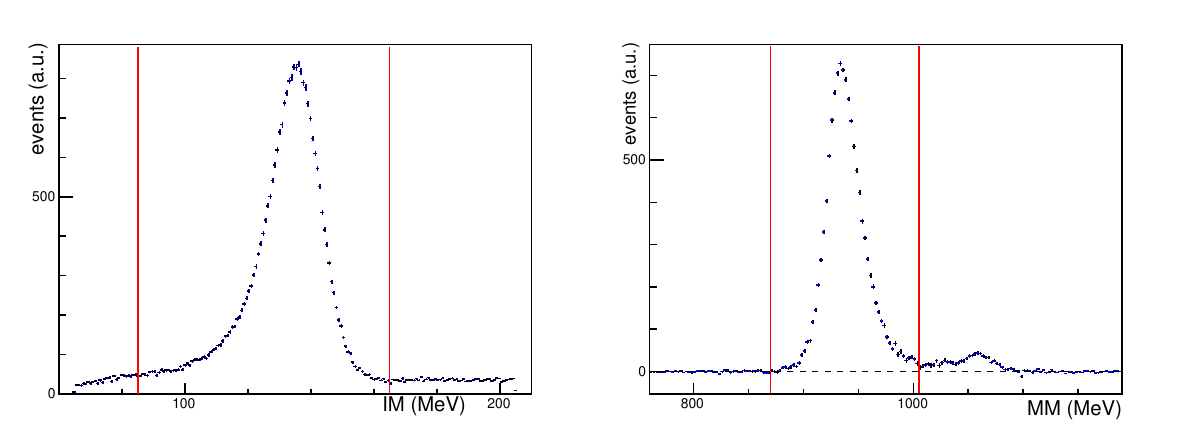}
  \caption{Left panel: two photon invariant mass (IM) distribution from the 
  $\pi^{0}$ reconstruction procedure obtained for the difference of 
  the candidate events with the parallel and antiparallel spin configurations. 
    Right panel: same as before, but for the missing mass (MM)
    distribution after the IM-based selection shown in the left panel. 
    In both cases, the acceptance region for the selection of a good $p\pi^0$ event is inside
    the vertical red lines.
    %
    %
  }
  \label{figure_IM}
\end{figure*}

The overall IM distribution obtained for the difference of
candidate events with the parallel and antiparallel spin configurations
is shown in the left panel of Fig.~\ref{figure_IM}, together with the cut ($\hbox{IM}\in[85-165]$~MeV) 
applied in the offline analysis that selected the events for further analyses.

The next step was to evaluate the missing mass (MM) of the event, where the recoil nucleon of the
reaction $\gamma  N \rightarrow \pi^{0} N$ was considered as a missing particle, even when it was detected.
%
The missing mass was calculated as follows:
\begin{equation}
  {\rm MM}=\sqrt{(E_{\gamma}+m_{p}-E_{\pi^{0}})^{2}-(\vec{p}_{\gamma}-\vec{p}_{\pi^{0}})^{2}} \ \ ,
  \label{MM_equation}
\end{equation}
where $E_\gamma$ and $\vec{p}_\gamma$ are the energy and momentum of the incoming photon in the lab frame, respectively,
$m_p$ is the proton mass in the initial state,
$E_{\pi^0}$ and  $\vec{p}_{\pi^0}$ are the reconstructed $\pi^{0}$ total energy and momentum, respectively.

The MM distribution obtained for the difference of candidate events after the previous analysis steps with the parallel and anti-parallel spin configurations is shown in the right panel of Fig.~\ref{figure_IM}.
Events from the $\pi^0\pi^0$ channel, coming from the high energy part of the photon Bremsstrahlung spectrum, 
can be clearly seen as a small peak in the right part of the distribution.
They are  rejected by the applied selection cut ($\hbox{MM}\in[870-1005]$~MeV) shown in the right panel of Fig.~\ref{figure_IM}.

To evaluate the residual contamination remaining after this cut, a sample of
$\gamma p \to \pi^0 \pi^0 p$ events was generated and their signal in the detection apparatus was simulated using a GEANT-based Monte Carlo code~\cite{geant4}, that accurately modeled the geometry and composition of the detection apparatus and considered the
applied electronic thresholds. This analysis showed that the fraction of the
$\pi^0\pi^0$ events passing the MM cut is about  $2\cdot 10^{-3}$.
Therefore their contribution was completely neglected in the rest of
the analysis.

\begin{figure}%
\centering
\includegraphics[scale=0.45]{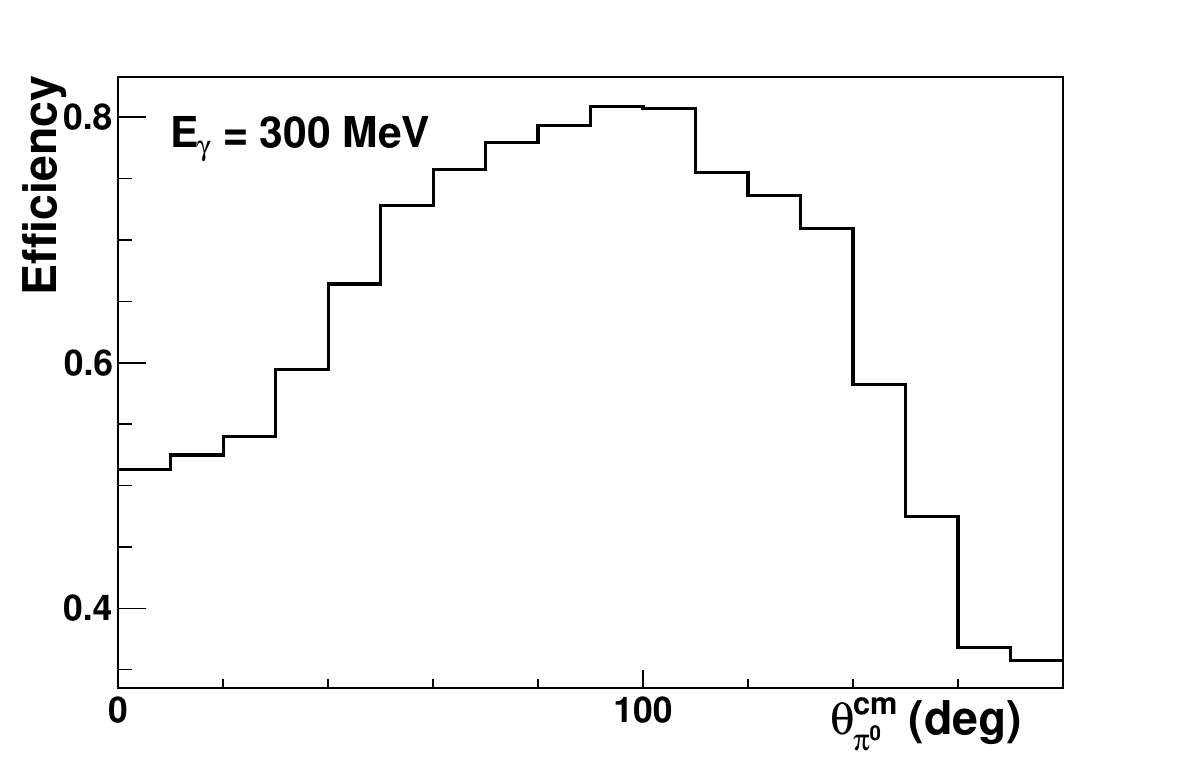}
\caption{$\pi^0$ reconstruction efficiency obtained using Monte Carlo simulated events
  at $E_\gamma=300$~MeV, as a function of the polar $\pi^0$ emission angle in
  the c.m.\@ system $\theta_{\pi^0}^{\text{cm}}$.}
\label{fig_geanteff}
\end{figure}

The detection and reconstruction efficiency of the $p\pi^0$ events
was also evaluated using the same GEANT-based code.
As an example, the simulated $\pi^0$ reconstruction efficiency
at the photon beam energy $E_\gamma=300$~MeV
is shown in Fig.~\ref{fig_geanteff} 
as a function the polar $\pi^0$ emission angle in the c.m. system ($\theta_{\pi^0}^{cm}$).
This  efficiency, over the full measured photon energy interval, varies in the forward, central, and backward angular regions within the ranges $[40\%-60\%]$,   $[50\%-90\%]$, and $[25\%-50\%]$, respectively.

The relative systematic uncertainty 
was evaluated by examining the different cuts and selection conditions applied to both
the experimental and the simulated data and it was estimated to be 10\% of the
value of the applied correction. 

%

 
\subsection{Systematic uncertainties} \label{par:syserr}

The various sources of systematic uncertainties discussed previously are summarized in Table~\ref{tab:syserr}.

Sources of common constant systematic uncertainties come from the photon flux normalization,
the beam and target polarization, and from the target surface density.
In contrast, the systematic uncertainty related to the $\pi^0$ reconstruction efficiency 
depends on both $E_\gamma$ and $\theta^{\text{cm}}_{\pi^0}$, and ranges
from $\sim$~1\% to $\sim$~7\% of the absolute $d\Delta\sigma/d\Omega$ values.

\begin{table}[ht]
\setlength{\tabcolsep}{12pt}
\centering{}
\caption{Relative systematic uncertainties given as total widths of uniformly-distributed values. \label{tab:syserr}}
\begin{ruledtabular}
\begin{tabular}{l c}
 \textbf{Source} & \textbf{Error} \\
  \hline
  Tagging efficiency & $\pm 4\%$ \\
  Beam polarization & $\pm 3\%$ \\
  Target polarization & $\pm 10\%$ \\
  Target density & $\pm 2\%$ \\
  $\pi^0$reconstruction efficiency & $\pm 1\%-7\%$ \\
%
\end{tabular}
\end{ruledtabular}
\end{table}

\section{Results}\label{sec:results}


\begin{figure*}%
\centering
\includegraphics[scale=0.4]{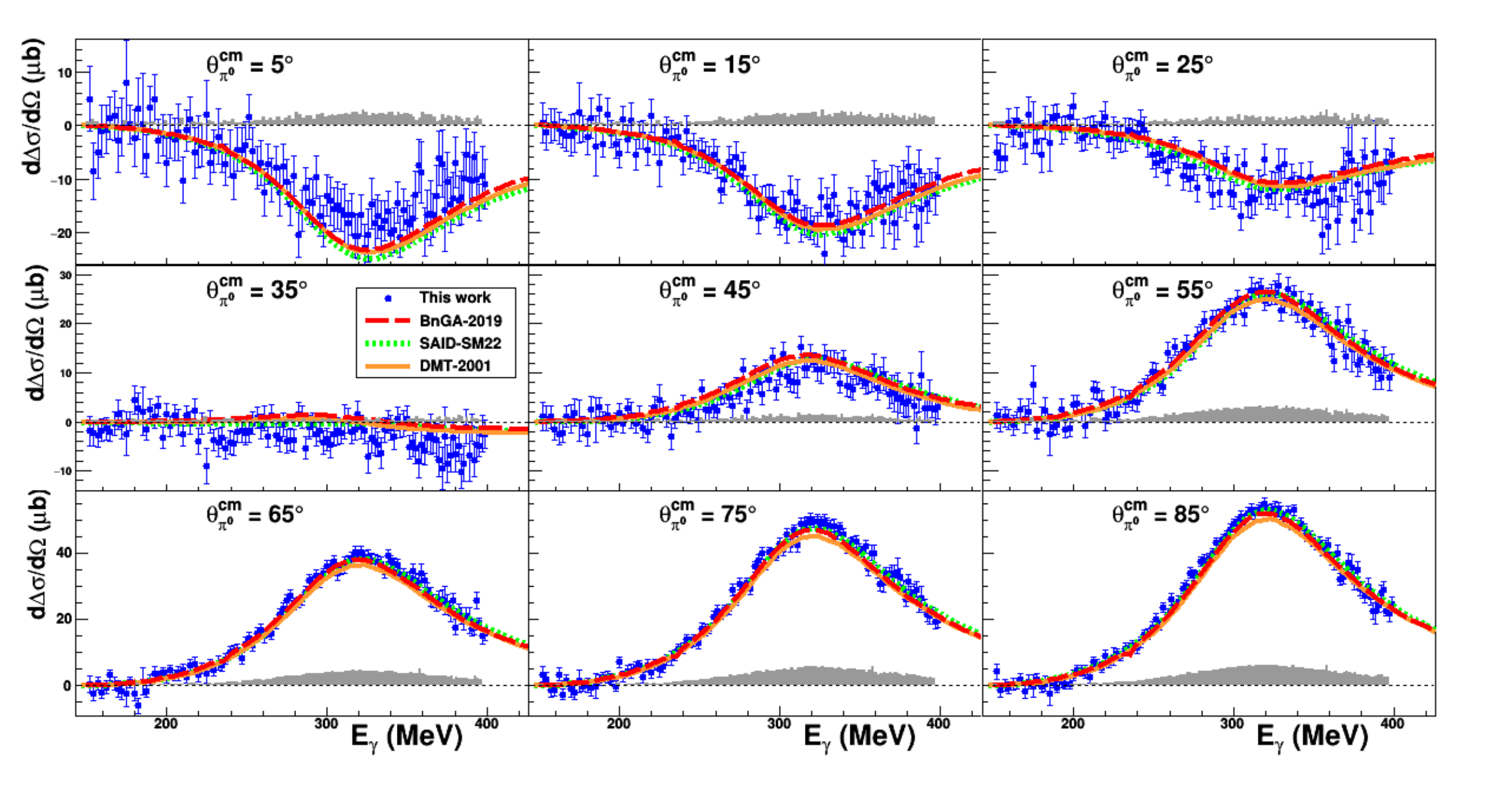}
\caption{
  Excitation functions of the
  $\vec{\gamma}\vec{p}\to p \pi^0$ reaction as
  a function of the photon beam energy for the measured forward $\theta_{\pi^0}^{\text{cm}}$
  bins. The experimental results in blue are compared with the predictions given by  the
  BnGa-2019 (dashed red lines)
  and SAID-SM22 (dotted green lines) energy-dependent PW analyses, and the DMT-2001 dynamical model (solid orange lines).
  The total contribution of all the systematic uncertainties,  given in half width units (see Sec.~\ref{par:syserr}),
  is shown as gray bars.
}
\label{fig_excit_1}
\end{figure*}

\begin{figure*}%
\centering
\includegraphics[scale=0.4]{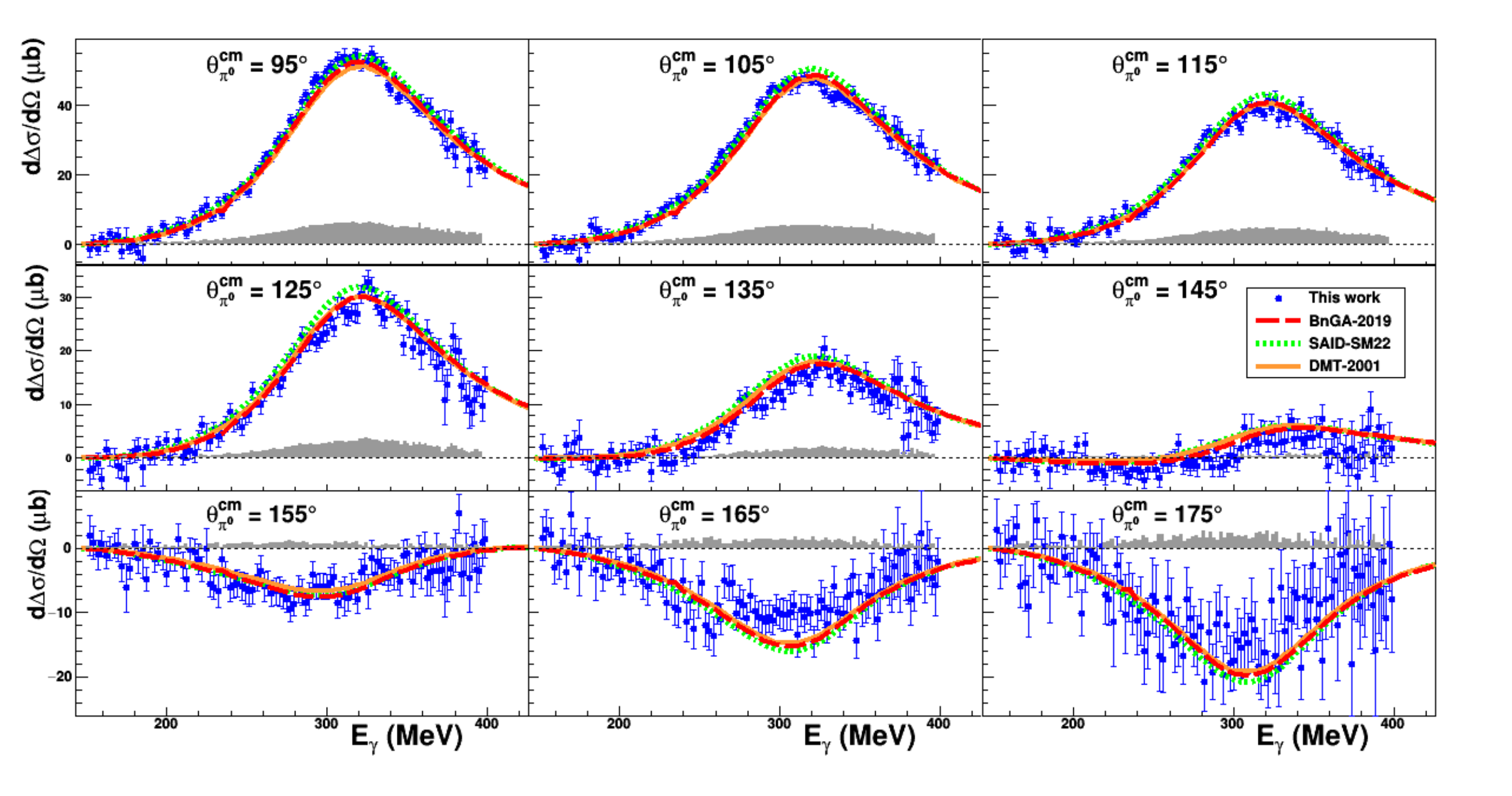}
\caption{
  Same as in Fig.~\ref{fig_excit_1}, but for
  for the measured backward $\theta_{\pi^0}^{\text{cm}}$ bins.
}
\label{fig_excit_2}
\end{figure*}

A compact representation of the differential cross section $d\Delta\sigma / d\Omega$ data
can be obtained by plotting them as excitation functions at the 
measured $\theta_{\pi^0}^{\text{cm}}$ angles, as shown in Figs.~\ref{fig_excit_1} and~\ref{fig_excit_2}.

The different lines show the predictions for this observable
given by two energy-dependent partial wave (PW) analyses:
BnGa-2019~\cite{boga} (dashed red lines),
SAID-SM22~\cite{SM22} (dotted green lines),
and by the DMT-2001 dynamical model~\cite{DMT0,DMT1,DMTX,DMT2} (solid orange lines).
The DMT-2001 model is a field-theoretical meson-exchange model for $\pi N$ scattering, optimized for the description of the single pion photoproduction process below $400$~MeV.
In contrast to the previous PW analyses,
 the energy dependence of
the DMT amplitudes is largely determined from theoretical considerations even
though there are free parameters, describing the non-resonant part of the amplitudes, that are determined from the fit of $\gamma N \to \pi N$ data.

In the comparison with our data, all the predictions show a rather good agreement
in all energy ranges,
except for slight differences at the very forward and backward angles.
\begin{figure*}%
\centering
\includegraphics[scale=0.75]{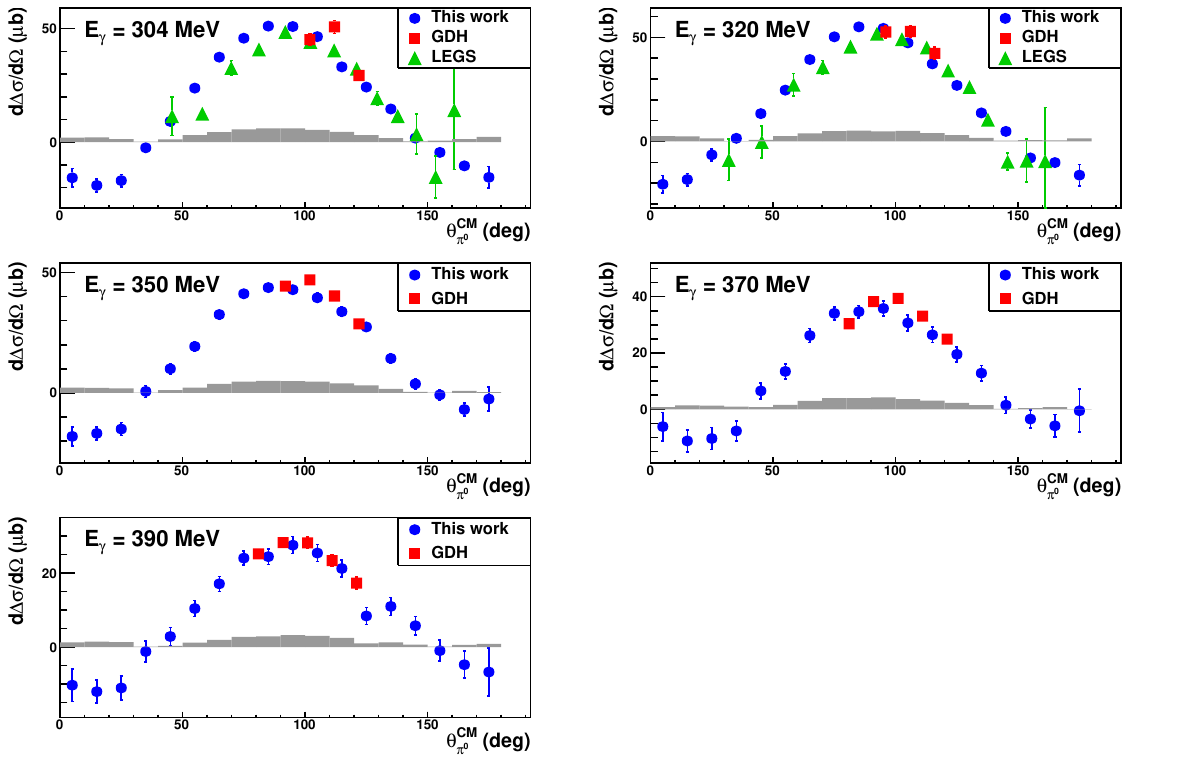}
\caption{
  The new data on the differential helicity-dependent cross section (blue circles) are compared
  with available published data  from the GDH~\cite{ahr04} (red squares) 
  and LEGS~\cite{sandlegs} (green up triangles) collaborations.
   The total contribution of the systematic uncertainties of the present data, given in half width units (see Sec.~\ref{par:syserr}),
  is shown as gray bars.
}
\label{fig_conf_gdhlegs}
\end{figure*}

\begin{figure*}%
\centering
\includegraphics[scale=0.7]{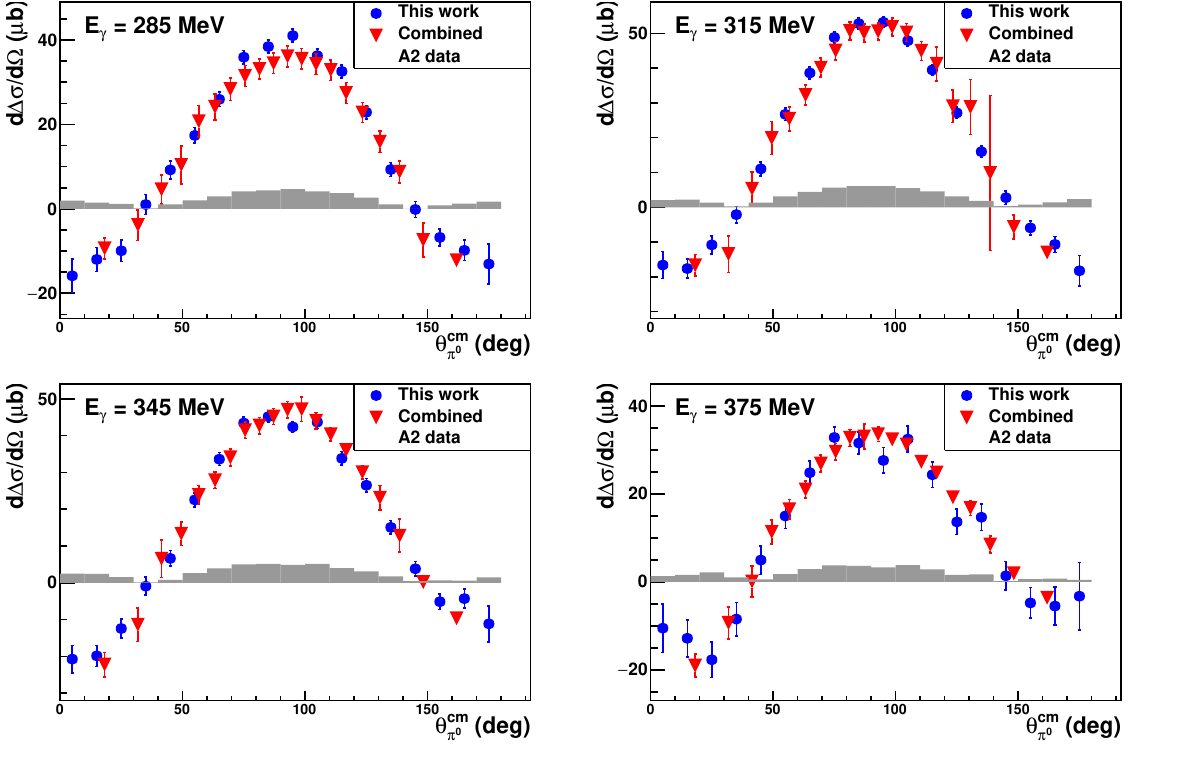}
\caption{
  The new data on the differential helicity-dependent cross section (blue circles)
  are compared to a combination of previously
  published A2 data~\cite{adlarson,farah} (red down triangles). See text for details.
   The total contribution of the systematic uncertainties of the present data, given in half width units (see Sec.~\ref{par:syserr}),
  is shown as gray bars.
}
\label{fig_conf_a2}
\end{figure*}

In Fig.~\ref{fig_conf_gdhlegs} 
some of the new  $d\Delta\sigma / d\Omega$ data are
compared, at fixed $E_\gamma$ values, to all available results for this observable, that were published by 
the GDH~\cite{ahr04} (red squares) and LEGS~\cite{sandlegs} (green up triangles)
collaborations.
As can be easily seen from the previous figures,
the present work, with 18 ($10^\circ$~wide) $\theta_{\pi^0}^{\text{cm}}$ bins 
measured at 114 $E_\gamma$ ($\sim\! 2$~MeV wide) bins,
provides a huge improvement in both quantity and quality
compared to the previous data in the $\Delta(1232)$ resonance region. 

As a further check, Fig.~\ref{fig_conf_a2} compares our results with 
a combination of previously  published data from the A2 Collaboration 
(red down triangles).
These combined values were obtained, for  $E_\gamma < 400$~MeV, by multiplying
the unpolarized differential cross section values $d\sigma_0/d\Omega$ from Ref.~\cite{adlarson}
with the recently published $E$ asymmetry values from Ref.~\cite{farah}, according to
the well-known identity (see, for instance, Ref.~\cite{wund}):
\begin{equation}
     \frac{d\Delta\sigma}{d\Omega} = -2 \cdot\frac{d\sigma_0}{d\Omega} E \ .
\end{equation}
The excellent agreement found with both the previously published GDH and  A2 data gives a strong indication
of the overall correctness of the offline analysis procedures.

%
\begin{figure*}%
\centering
\includegraphics[scale=0.35]{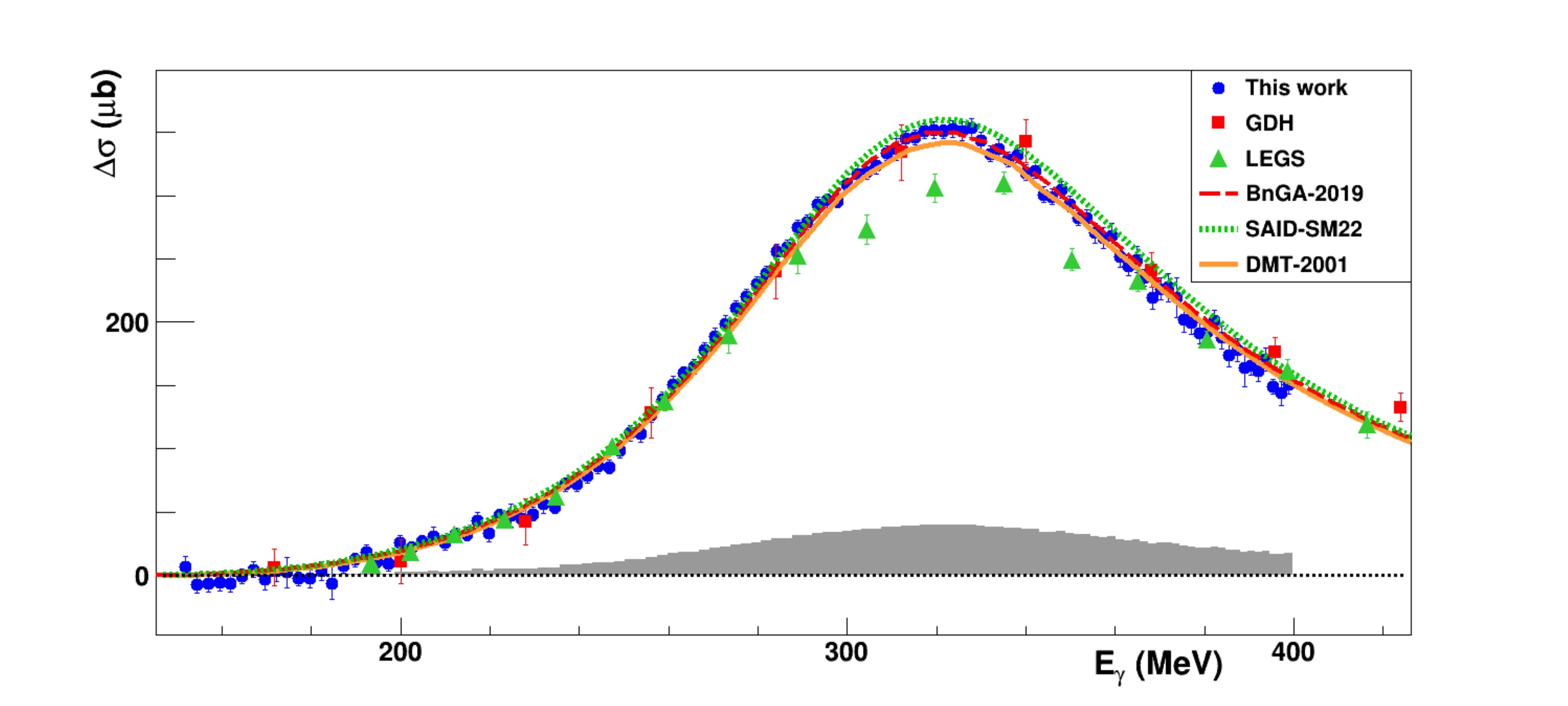}
\caption{The new total helicity-dependent cross section $\Delta\sigma$
  for the  $\gamma p \to p \pi^{0}$ channel (blue circles) is compared with
  the results obtained by the GDH~\cite{ahr04} (red squares) and
  LEGS~\cite{sandlegs} (green up triangles)
  collaborations.
  The different line styles show the predictions of the  
BnGa-2019~\cite{boga} (dashed red line) 
 and  SAID-SM22~\cite{SM22} (green dotted line) PW analyses, and  of the DMT-2001 dynamical model~\cite{DMT0,DMT1,DMTX,DMT2} (solid orange line).
  The total contribution of the systematic uncertainties, given in half width units (see Sec.~\ref{par:syserr}),
  is shown as gray bars.
}
\label{cstot}
\end{figure*}

The total helicity-dependent cross section $\Delta\sigma$ 
for the $\gamma p \to p \pi^{0} $ reaction, obtained by integrating
the differential cross section
 $d\Delta\sigma / d\Omega$ 
over the full solid angle,
is shown in Fig.~\ref{cstot} 
(blue points) for the photon beam energy range from $E_\gamma$=150~MeV up to 400~MeV.
It is again compared to the predictions given by the previous PW analyses and model and to
the  avaliable data  from the GDH~\cite{ahr04} (red points) and LEGS~\cite{sandlegs} (green points)
collaborations.

As already noticed for the excitation functions shown in Figs.~\ref{fig_excit_1} and \ref{fig_excit_2},
our data are in excellent agreement with the BnGa-2019 and SAID-SM22 PW analyses, the DMT-2001 model,
and with the GDH data,  while differing from the
LEGS data in the region around the $\Delta$ resonance mass.
As can be clearly deduced from the differential cross section data shown in Fig.~\ref{fig_conf_gdhlegs},
this discrepancy comes from 
differences in the angular shape,
especially present 
in the forward $\theta_{\pi^0}^{\text{cm}}$ region, that can not be accommodated with a single common scale shift.

\section{Legendre analysis of the $\bm{d\Delta\sigma/d\Omega}$ data} \label{sec:legendre_analysis} 

The multipole content of the reaction amplitude can be easily accessed by 
expanding the measured $d\Delta\sigma/d\Omega$ angular distributions in
a (truncated) Legendre series.

The energy dependence of the expansion coefficients
can reveal specific correlations between individual resonance states of
certain parities (see, for instance, Ref.~\cite{wund} and references therein).
This method proved to be particularly effective for $E_\gamma < 400$~MeV, where
only one well-known resonance, the $\Delta(1232)$, dominates 
the amplitude and low-lying multipoles
can be accessed via the interference terms with the
dominant $M_{1+}$ term.  

The Legendre coefficients $a_k$ were then obtained by
fitting a series of  associated Legendre polynomials $P_k$ to the angular distributions  $d\Delta\sigma/d\Omega$ with the function (see Ref.~\cite{wund}):
\begin{equation}
  \frac{d\Delta\sigma}{d\Omega} = -2\,\frac{q}{k}\cdot
{\displaystyle \sum_{k=0}^{2\ell_{\text{max}}}(a_{\ell_{\text{max}}})_k({W})P_{k}(\cos\theta)\ ,}
\label{Leg_fit}
\end{equation}
where $q, k$, and $W$ are the photon and $\pi^0$ momenta, and  the total energy in the c.m. system, respectively.
The notation $(a_{\ell_{\text{max}}})_{k}$ means that in the fitting procedure only the partial waves with the $\pi N$ relative angular momentum up to $\ell=\ell_{\text{max}}$ were included in the fit.
The multipoles contributing to the fit for $\ell_{\text{max}}=1,2$
are listed in Table~\ref{tab:Sensibility-on-multipoles}.  

\begin{table}[ht]
\centering{}
\setlength{\tabcolsep}{12pt}
\caption{The multipole amplitudes contributing to the fitted cross section
  reported in Eq.~(\ref{Leg_fit})
  for different choices of $\ell_{\text{max}}$.  \label{tab:Sensibility-on-multipoles}}
\begin{ruledtabular}
\begin{tabular}{ccc}
\textbf{$\mathbf{\ell_{\text{max}}}$} & \textbf{Wave} & \textbf{M-poles}\tabularnewline
\hline
\multirow{2 }{* }{1} & $s$-wave & $E_{0+}$
\tabularnewline
 & $p$-wave & $E_{1+},M_{1+},M_{1-}$\tabularnewline
2 & $d$-wave & $E_{2+},E_{2-},M_{2+},M_{2-}$\tabularnewline
\end{tabular}
\end{ruledtabular}
\end{table}

%
\subsection{Bootstrap-based fitting procedure}\label{sec:fit}

As noted in Ref.~\cite{wund}, 
an important issue when using this type of algorithm is the proper handling of
the systematic uncertainties associated with the data being fit.
A widely used method, when one single common and Gaussian-distributed
multiplicative systematic uncertainty is present, is to introduce a modified $\chi^2$ function,
with an additional overall scale parameter to be fitted (see, for instance, Refs.~\cite{DAgostini1994,Pedroni:2019dlg}):
\begin{equation}
 \label{eq:chi2sys}
 \chi_{\text{mod}}^2  (\bm\vartheta) = \sum_i{\left(\frac{fy_i - T_i(\bm\vartheta)}{f\sigma_i}\right)^2}
 +\left(\frac{f-1}{\sigma_{sys}}\right)^2.
 \end{equation}
Here
$y_i$ and $\sigma_i$ are the experimental values to be fitted and
their corresponding statistical uncertainties in root mean square units (rms), respectively. 
$T_i$ are the theoretical predictions given by a model depending
on the set of unknown parameters $\bm\vartheta$ to be evaluated from the data, 
$f$ and $\sigma_{sys}$ are the additional fit parameter and its estimated uncertainty (in rms units), respectively.

However, Eq.~(\ref{eq:chi2sys}) cannot be used to fit the new $d\Delta\sigma/d\Omega$ data, 
since they have both uniform and angular-dependent systematic uncertainties, as discussed in Sec.~\ref{par:syserr}, and these effects can not be properly accounted for by the single factor $f$.
Moreover, under these conditions, the goodness-of-fit distribution
is generally not given by the $\chi^2$ function and the fit parameter
errors are not a priori Gaussian, since the sum appearing in
Eq.~(\ref{eq:chi2sys}) consists of correlated and non-Gaussian variables.

To overcome all these difficulties,
the fit to the present data was performed by using
an innovative method~\cite{Pedroni:2019dlg}, based on the parametric bootstrap technique, already
successfully deployed for different analyses of nuclear
Compton scattering data (see Refs.~\cite{Pasquini:2018,Pasquini:2019nnx, Mornacchi:2022}).

As an example, in the case of a single data set with
Gaussian statistical errors and uniform common multiplicative systematic
uncertainties, this method consists of randomly generating 
$N$ Monte Carlo replicas of the experimental
data, where each data point $y_{i}$ is replaced by:
\[
y_{i} \rightarrow y_{i}^{(b)} = (1 + \delta_{b})(y_{i} + r_{i,b} \sigma_{i}) \ .
\]
The indices $i$ and $b$ run over the number of data points
and the bootstrap replica,
respectively; $r_{i,b}$ is a random number extracted from the normal distribution $\mathcal{N}(0,1)$, 
and $\delta_{b}$ is a uniform random variable distributed according to $\mathcal{U}[-\Delta,\Delta]$,
where $\pm\Delta$ is the quoted systematic uncertainty interval.
A set of fitted parameters (${\bm\vartheta}_{b}$) is extracted from each of the $N$ simulated replicas and,
at the end of the procedure, the best values 
($\hat{\bm\vartheta}_{\text{boot}}$)
and the errors of the fitted quantities 
can be easily extracted from the empirical statistical parameters of the
 resulting distributions.

The goodness of fit (see again Ref.~\cite{Pedroni:2019dlg}) is evaluated by comparing the minimum $\hat{\chi}_{\text{boot}}^2$
value obtained as 
\begin{equation}
 \label{eq:chi2boot}
\hat{\chi}_{\text{boot}}^2  = \sum_i{\left(
\frac{y_i - T_i(\hat{\bm\vartheta}_{\text{boot}})}{\sigma_i}
\right)^2 } \ ,
 \end{equation}
with the theoretical distribution empirically obtained
by repeating the bootstrapping algorithm with each experimental datum
$y_i$ replaced by  $T_i(\hat{\bm\vartheta}_{\text{boot}})$.
With this procedure, after 
a suitably large number $k$ of
bootstrap cycles, an estimate of the
goodness-of-fit probability function can be 
obtained from the set of minimum values
\[
\hat{\chi}_{th,\text{boot},1}^2, \hat{\chi}_{th,\text{boot},2}^2, \ldots,
\hat{\chi}_{th,\text{boot},k}^2  
\]
evaluated at the end of each cycle. This ensures that the correct $p$-value is then always provided by the present fitting procedure.


%
\subsection{Fit results}

A total of $N=10^4$ bootstrap samples were generated for each $W$ bin,
and the minimization was performed at the end of each iteration.

All the different systematic uncertainties were assumed
to follow a uniform distribution
over the interval defined by the bounds given in Table~\ref{tab:syserr}.
In the case of the angular-dependent source,
the same common fraction of the full variation interval was randomly generated,
for all angular bins,
at each bootstrap cycle.
The final total uncertainty factor entered into the bootstrap procedure
is given by the product of all the random uniform variables generated by the previous procedure.
\begin{figure}%
\centering
\includegraphics[scale=0.16]{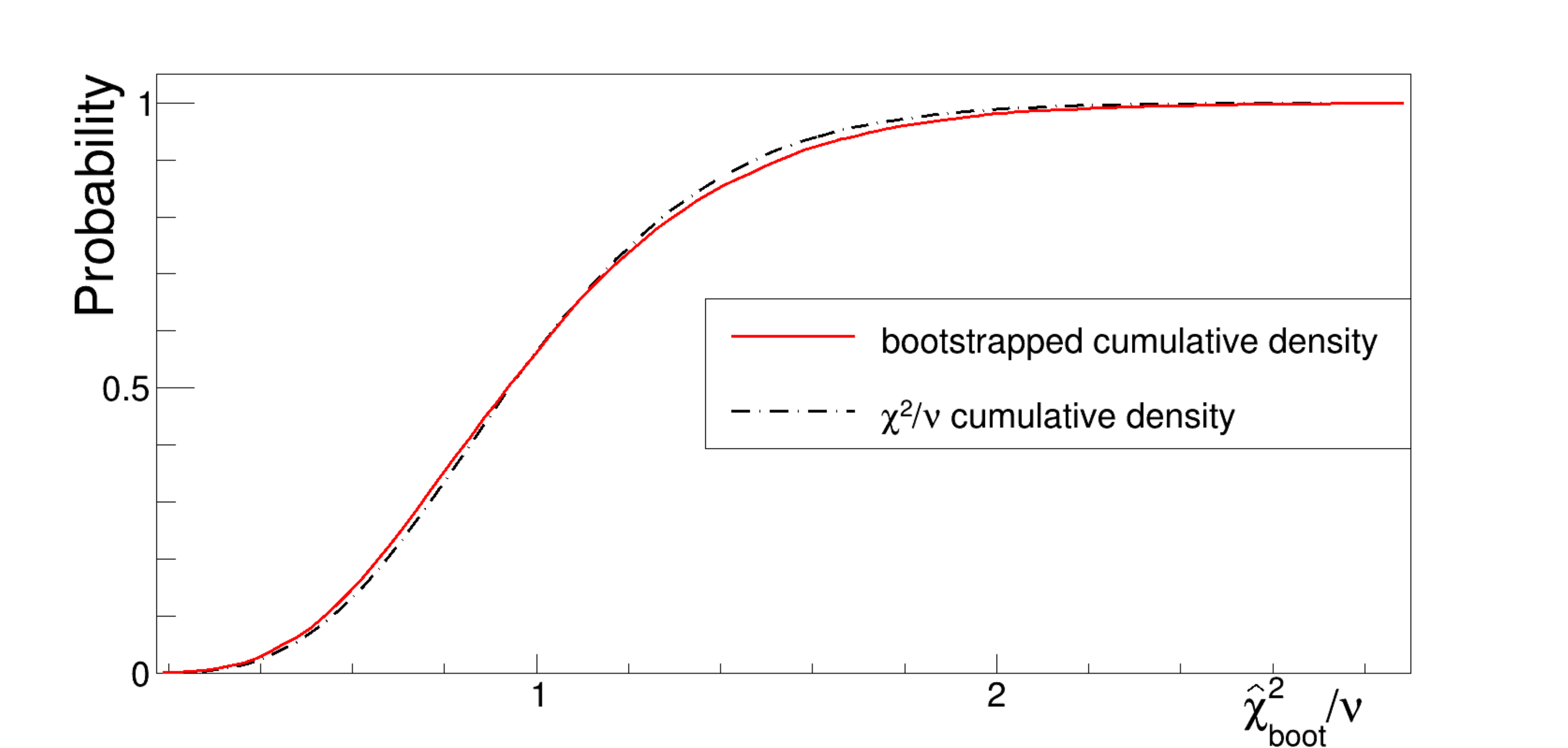}
\caption{The cumulative density function (CDF) of the goodness-of-fit distribution divided by the  number of
  degrees of freedom ($\nu$), as obtained from the bootstrap
  procedure (red solid line) at $E_\gamma=350$~MeV, is compared with the CDF of the reduced $\chi^2$ distribution (black dash-dotted line). }
\label{fig_chi2_1}
\end{figure}

\begin{figure*}%
\centering
\includegraphics[scale=0.3]{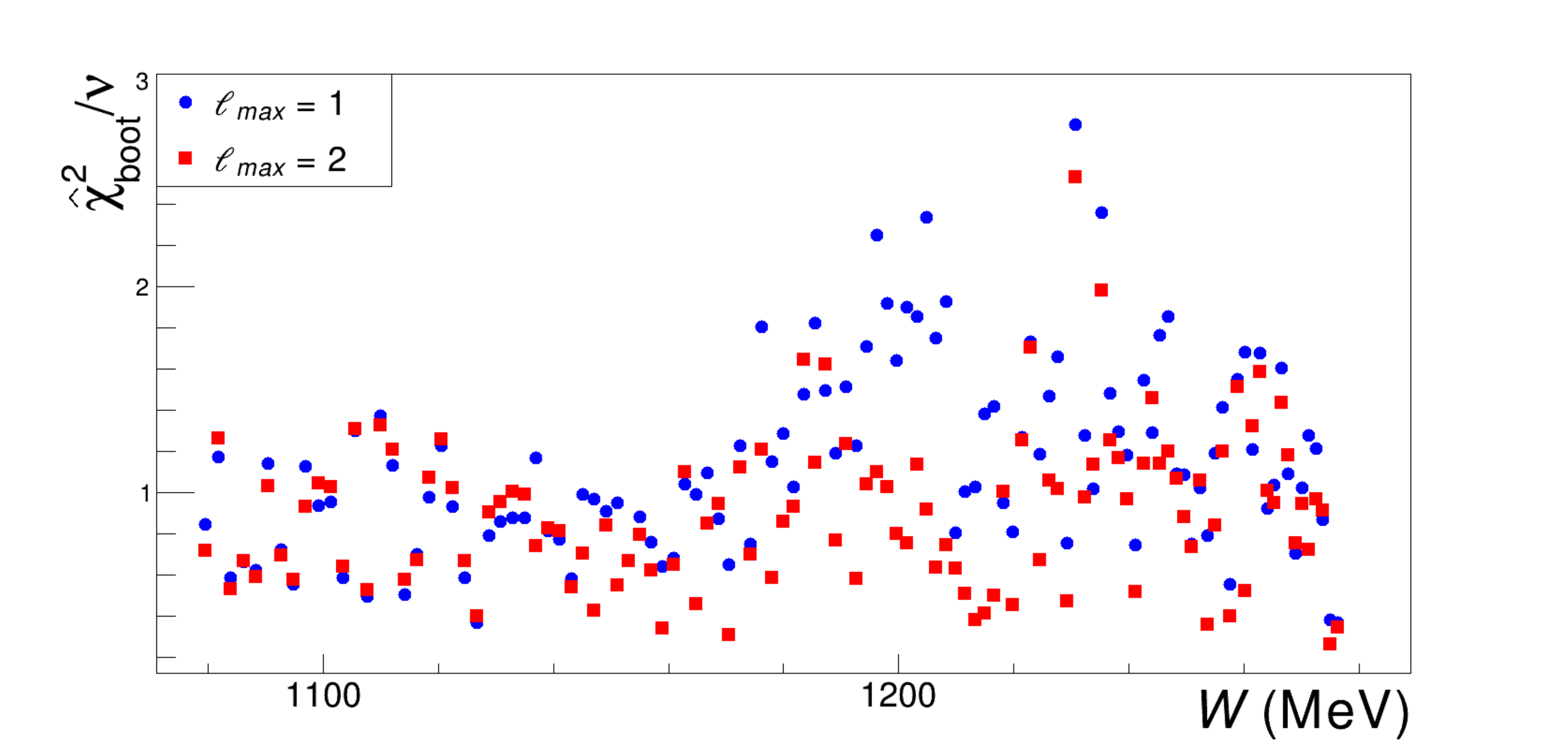}
\caption{Reduced $\hat{\chi^2}$ values obtained from the bootstrap procedure, as a function of the center-of-mass energy $W$ at  $\ell_{\text{max}}=1$ (blue circles) and 
 $\ell_{\text{max}}=2$ (red squares).
}

\label{fig_chi2_2}
\end{figure*}

First, we evaluated the expected goodness-of-fit distribution,
from which the $p$-value associated with the minimum $\hat{\chi}_{\text{boot}}^2$
values obtained by the bootstrap procedure has to be computed.
Figure~\ref{fig_chi2_1} shows the obtained cumulative density function (CDF)
divided by the number of degrees of freedom ($\nu$), 
obtained at $W=1240$~MeV ($E_\gamma = 350$~MeV) and with $\ell_{\text{max}}=1$,
compared with the CDF of the reduced $\chi^2$ distribution.
These two distributions were found to basically coincide, and 
the same result was obtained for all  $W$ bins as well as for $\ell_{\text{max}}=2$.


In the second step of this procedure,
we  evaluated the minimum value of the $\ell_{\text{max}}$ parameter 
that could reproduce sufficiently well the behavior of the
experimental data.
This was done by comparing, at each $W$ bin,
the $\hat{\chi}^2_{\text{boot}}$ values obtained for both
$\ell_{\text{max}} =1$ and $\ell_{\text{max}} =2$
after a simultaneous fit of all the involved coefficients, as shown in in Fig.~\ref{fig_chi2_2}. 

Except for a few points in the center-of-mass energy region $W\sim 1200$~MeV, the 
fit with $\ell_{max}=2$ gives, in general, no significant improvement in the
$\hat{\chi}^2_{\text{boot}}$ values. 
In several cases where the difference is significant, the $\hat{\chi}^2_{\text{boot}}$ values obtained
with $\ell_{max}=2$ have suspiciously low values which may indicate overfitting.

\begin{figure*}%
\centering
\includegraphics[scale=0.32]{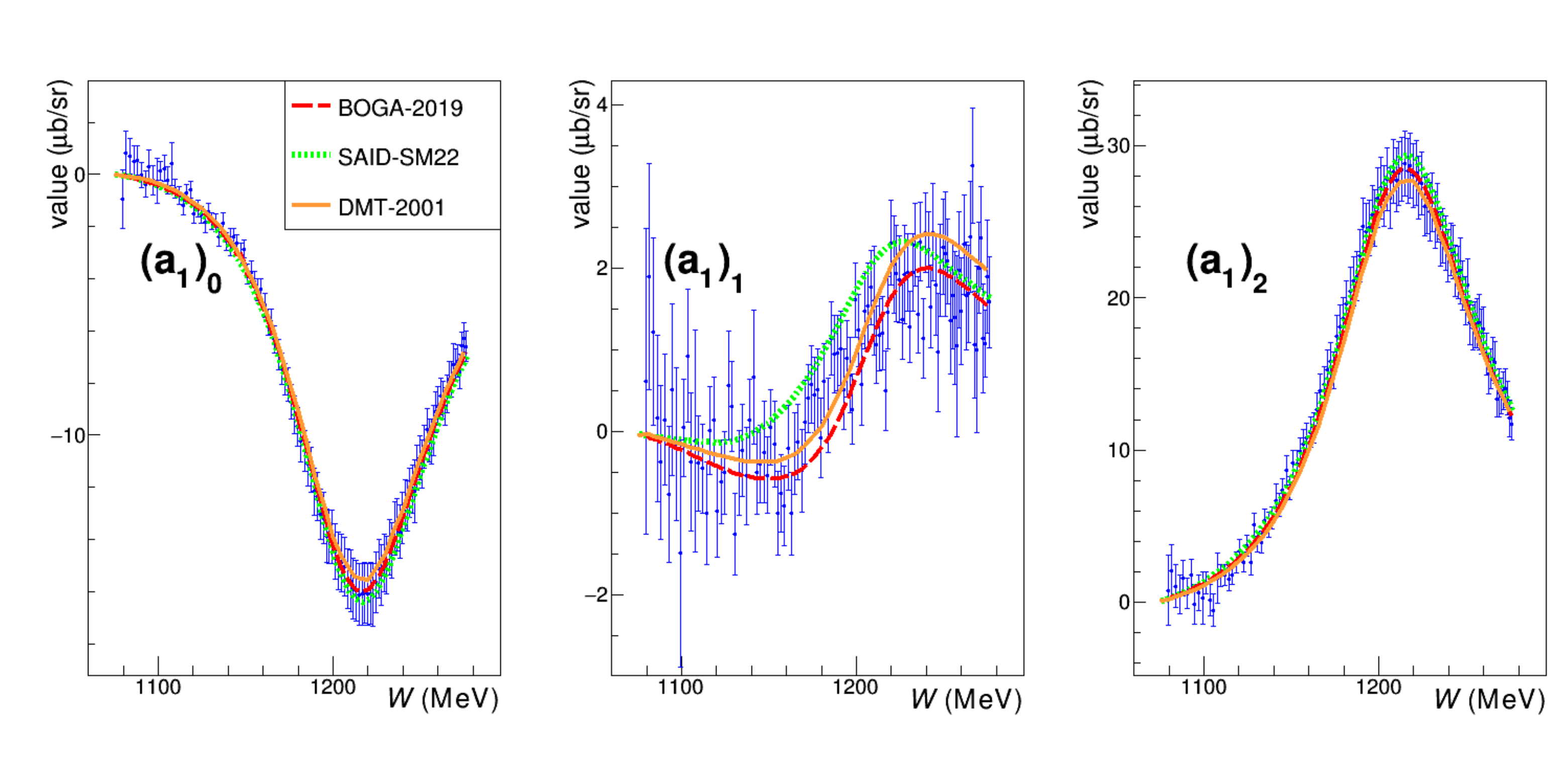}
\caption{The fitted Legendre coefficients $(a_1)_0$, $(a_1)_1$, and $(a_1)_2$ are compared
with the predictions of the 
BnGa-2019~\cite{boga} (dashed red lines) and
SAID-SM22~\cite{SM22} (dotted green lines)
PW analyses, and of
the DMT-2001 dynamical model~\cite{DMT0,DMT1,DMTX,DMT2} (solid orange lines).
}
\label{fig_leg}
\end{figure*}

In addition, for $\ell_{max}=1$, we have that, after averaging
over all $W$ bins, $\vmed{\hat{\chi}_{\text{boot}}^2/\nu} = 1.14$,  which corresponds to a $p-$value of about $31\%$.
All previous indications lead us to conclude that the fit with $\ell_{\text{max}}=1$
reproduces our data sufficiently well.
Such a result was to be expected since, 
in the $\Delta(1232)$ region, the production of
$s$ and $p$-wave pions gives, by far,
the largest contribution to the $\pi^0$ channel, while $d$-waves only 
contribute thanks to very small  interference terms with the dominant $p$-waves.
Their effect can be quantified by the value of the $(a_2)_3$ Legendre coefficient
calculated with the previous PW analyses and model. 
In the $\Delta(1232)$ region, the absolute value of this coefficient 
was found to be $\lesssim 1 \mu b/sr$, 
more than one order of magnitude smaller than the absolute contributions
given by the $(a_1)_0$ and $(a_1)_2$ coefficients, dominated by the $p$-waves
multipoles, and also at least 2 times lower than the
$(a_1)_1$ coefficient, which quantifies the small
 $s$-$p$ interference contribution (see Fig.~\ref{fig_leg}).
 Contributions due to $d$-wave pions can then be safely neglected in the present context.

\begin{figure*}%
\centering
\includegraphics[scale=0.75]{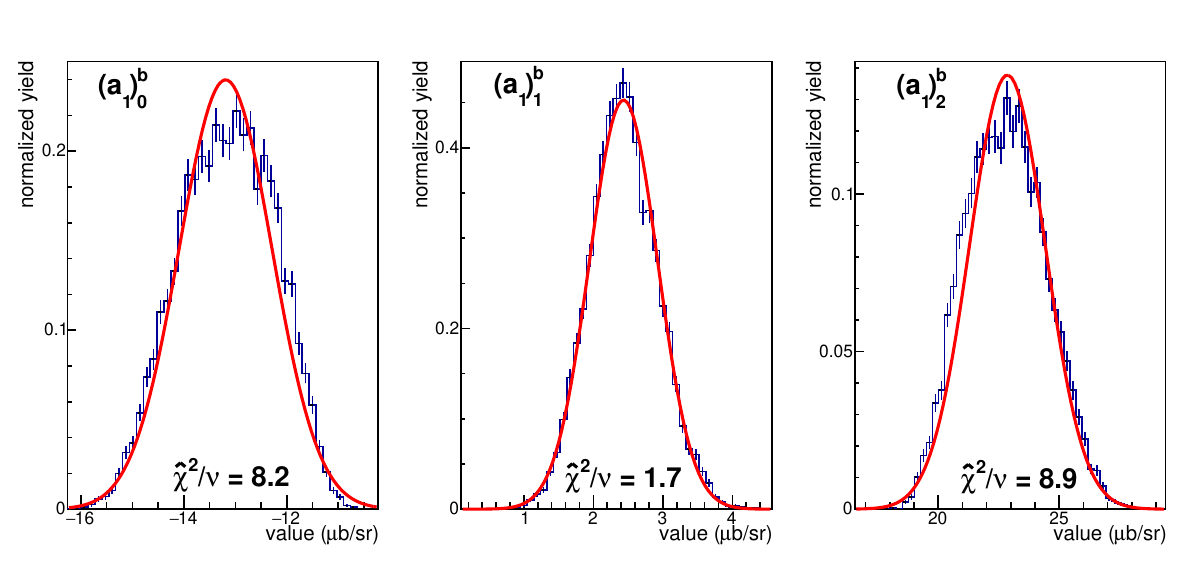}
\caption{Probability distributions for the fit parameters $(a_1)^b_0$, $(a_1)^b_1$, and $(a_1)^b_2$
as determined at the end of each bootstrap replica at $W=1240$~MeV ($E_\gamma = 350$~MeV). 
These distributions are compared with the best-fit Gaussian curves (red lines),
and the corresponding  $\hat{\chi}^2/\nu$ values are also given at the bottom of each plot.
}
\label{fig_coeff}
\end{figure*}

An example of the probability distributions for the fit
parameters $(a_1)^b_0$, $(a_1)^b_1$, and $(a_1)^b_2$, computed at the end of each
bootstrap replica, is given in Fig.~\ref{fig_coeff}, where
these densities are plotted at $W=1240$~MeV ($E_\gamma=350$~MeV).   

For the reasons discussed earlier, 
these distributions, especially for the $(a_1)^b_0$ and $(a_1)^b_2$ parameters,  show significant deviations from the pure Gaussian shape, as indicated by the comparison with the best-fit Gaussian red curves. 
%
Very similar density functions were found for all other $W$ bins.

The values of the Legendre coefficients $(a_1)_0$, $(a_1)_1$, and $(a_1)_2$,
determined as the mean of distributions obtained by the bootstrap-based fit procedure at each $W$ bin,
are plotted in Fig.~\ref{fig_leg}.
The quoted fit errors are the 68\% confidence level (CL) determined using the quantiles
of the bootstrapped parameter distributions, 
and include the contribution of both the statistical and systematic uncertainties of the experimental data. 
The numerical values of all these coefficients are reported in
  Table~\ref{tabref} of the Appendix.

The different curves represent the corresponding coefficients
evaluated with the BnGa-2019 and 
SAID-SM22 PW analyses with the DMT-2001 dynamical model.
A quite good agreement is found with 
all the previous predictions for the $(a_1)_0$ and  $(a_1)_2$ coefficients, whose value is mostly determined by the dominant $p$-wave contributions (see later Eqs.~\ref{eq:leg0}-\ref{eq:leg1}). On the contrary, these predictions give sizeable differences  in the value of the $(a_1)_1$ coefficient, which quantifies the effects of the  $s-p$ interference terms.
Our new data will allow to resolve these discrepancies and improve our understanding of the role of the low-lying multipoles for the $\pi N$ process in the $\Delta(1232)$ energy region.  

%
 

\newpage

\section{Derivation of the E2/M1 ratio}\label{sec:e2m1}

In the $\ell_{max}=1$ approximation (i.e., only $s$ and $p$-waves are considered - see Table~\ref{tab:Sensibility-on-multipoles}), 
the Legendre coefficients of the helicity-dependent differential cross section can be decomposed
in terms of multipoles as~\cite{wund,wund2}:

\begin{widetext}
\begin{equation}\label{eq:leg0}
  \begin{aligned}
    (a_1)_0 &= |E_{0+}|^2 + |M_{1-}|^2 +3E^*_{1+} (3E_{1+} + M_{1+}) +
      M^*_{1+}(3E_{1+} - M_{1+}) \ ; \\
    (a_1)_1 &{}= E^*_{0+} (3E_{1+} - M_{1-} + M_{1+}) +
     E_{0+}(3E^*_{1+}- M^*_{1-} +M^*_{1+}) \ ; \\
    (a_1)_2 &=  -M^*_{1-} (3E_{1+} + M_{1+}) +
     E^*_{1+} (6E_{1+} - 3M_{1-}) 
    - M^*_{1+}(M_{1-} - 2M_{1+}) \ .
      \end{aligned}
\end{equation}
\end{widetext}

To evaluate $R_{EM}$ (see Eq.~(\ref{eq:1new})) from these equations, the following two approximations were sequentially applied:
\begin{itemize}
\item[(i)] all terms involving only the $E_{0+}$, $E_{1+}$, and $M_{1-}$ multipoles were dropped
as, in the region of the $\Delta(1232)$ resonance mass,
they can be assumed to be negligible compared to the terms where $M_{1+}$ contributes;
\item[(ii)] a perfect Breit-Wigner form was assumed for the $M_{1+}$ multipole,
which gives, by far, the largest contribution
to the $\Delta(1232)$ resonance excitation.
Under this assumption, the real part of the $M_{1+}$
multipole vanishes exactly at the $M_\Delta$ value.
%
We can therefore set:
\[
M_{1+} \simeq i \hbox{Im}[M_{1+}].
\]
\end{itemize}

The first step in deriving $R_{EM}$ was to apply approximation (i) to the coefficients given in Eq.~(\ref{eq:leg0}):
\begin{widetext}
\begin{equation}\label{eq:leg1}
    \begin{aligned}
  (a_1)_0 &\simeq - |M_{1+}|^2 +3E^*_{1+} M_{1+} + 3M^*_{1+}E_{1+}  
      = - |M_{1+}|^2  + 
        6\hbox{Re}[E^*_{1+}M_{1+}] \ ; \\
  (a_1)_1 &\simeq 2\hbox{Re}[E^*_{0+}M_{1+}] \ ;  \\
  (a_1)_2 &\simeq -2\hbox{Re}[M^*_{1-}M_{1+}]+ 2|M_{1+}|^2 \ .
   \end{aligned}
\end{equation}
\end{widetext}
It can be immediately noticed that the coefficient $(a_1)_1$ is irrelevant for further considerations
and will be ignored from now on.
Furthermore, applying approximation (ii) resulted in:
\begin{widetext}
\begin{equation}\label{eq:a2} 
 \begin{aligned}
  (a_1)_0 &\simeq  -\hbox{Im}[M_{1+}]^2+6\hbox{Im}[E_{1+}]\hbox{Im}[M_{1+}] \ ; \\
  (a_1)_2 &\simeq  2\hbox{Im}[M_{1+}]^2   -2\hbox{Im}[M_{1-}] \hbox{Im}[M_{1+}] \simeq 
   2\hbox{Im}[M_{1+}]^2 \ ,
 \end{aligned}
 \end{equation}
 \end{widetext}
where, in the last step of the previous equation,
it was considered that
$ \hbox{Im}[M_{1-}] \ll \hbox{Im}[M_{1+}]$
 around the $M_\Delta$ value.

Using Eq.~(\ref{eq:a2}), and recalling Eq.~(\ref{eq:1new}), the ratio $(a_1)_0/(a_1)_2$
can now be calculated as:
\begin{widetext}
\begin{equation}
  \frac{(a_1)_0}{(a_1)_2} \simeq  
  - \frac{\hbox{Im}[M_{1+}]^2}{2\hbox{Im}[M_{1+}]^2}+ 
   \frac{6\hbox{Im}[E_{1+}]\hbox{Im}[M_{1+}]}{2\hbox{Im}[M_{1+}]^2}=  
    3 R_{EM} - \frac{1}{2} \ . \label{eq:approx}
\end{equation}
\end{widetext}

The following quantity can then be defined as a suitable approximation for $R_{EM}$:
 \begin{equation}\label{eq:rem}
R_L =  \frac{1}{3} \frac{(a_1)_0}{(a_1)_2} + \frac{1}{6} \simeq R_{EM}\ .
 \end{equation}
This allows for the extraction of the parameter $R_L$ from the ratio
of only two Legendre coefficients fitted from the angular distribution of 
only one observable, $d\Delta\sigma/d\Omega$.
This procedure has two main advantages: i) it reduces by one the number of experimental observables
to be measured compared to some of the previous extractions (e.g., Ref.~\cite{beck3}); 
ii) the experimental uncertainty of $R_L$ is independent of all common scaling
systematic uncertainties affecting the experimental data, as previously noted in Ref.~\cite{beck2}.

\begin{figure*}%
\centering
\includegraphics[scale=0.325]{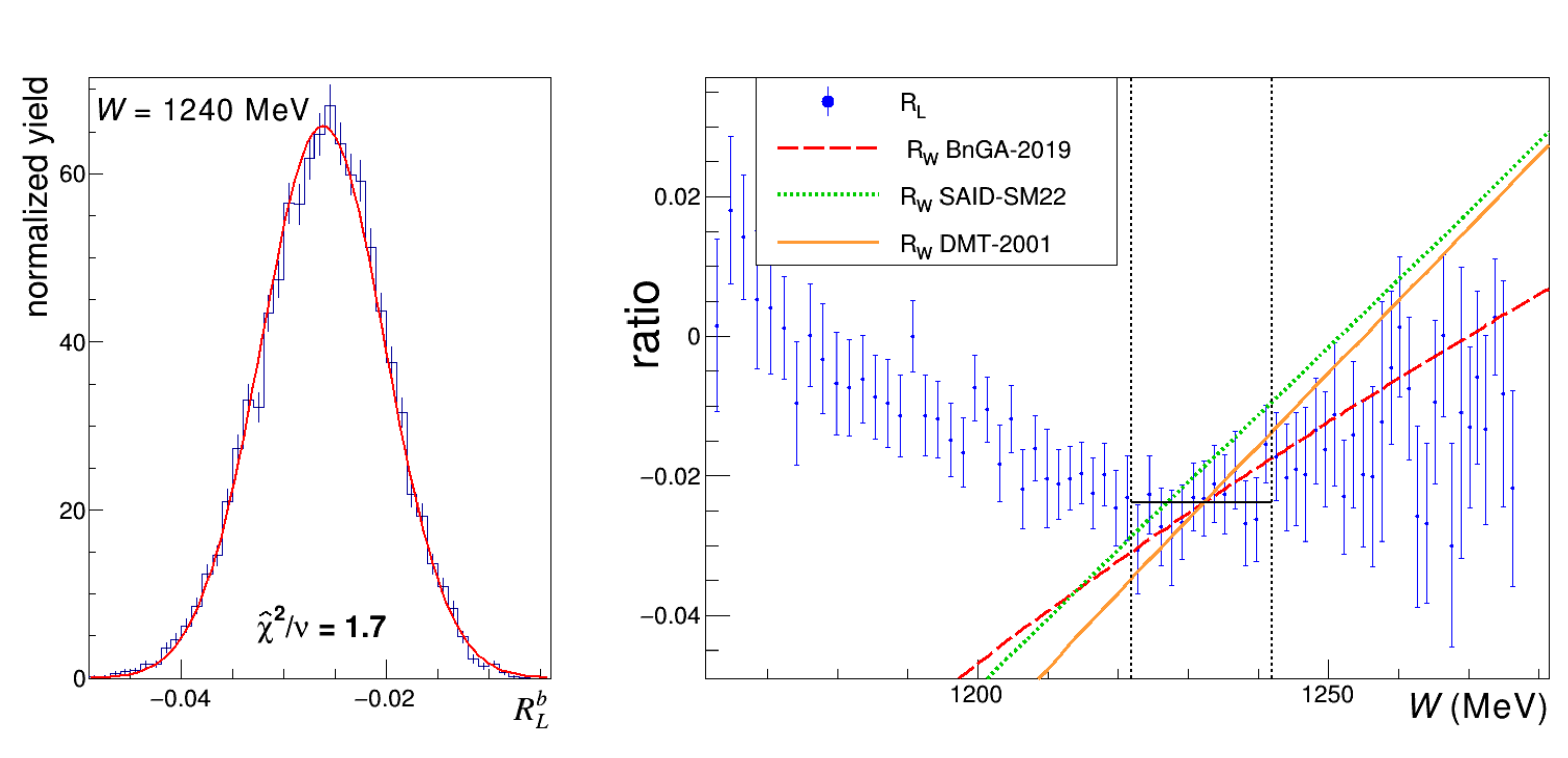}
\caption{ Left panel: probability distribution for the $R_b$ parameter (see Eq.~(\ref{eq:rem})),
  as determined by the bootstrap procedure at 
  $W = 1240$~MeV ($E_\gamma = 350$~MeV).
  This distribution is compared to the best-fit Gaussian curves (red lines),
  and the corresponding $\hat{\chi}^2/\nu$ value is also given in the canvas.
  Right panel:  The $W$ (total c.m. energy) dependence of $R_L$ (see eq.~\ref{eq:rem})
  for the upper part
  of the measured photon energy interval
  is compared to the $R_W$ ratio (see eq.~\ref{eq:rem-multi}) predicted by BOGA-2019 (dashed red line), SAID-SM22 (dotted green line) and DMT-2001 (solid orange line).
  The vertical dotted black lines
  define the $W$ region $1232\pm 10$~MeV selected to evaluate $R_{EM}$, while
  the horizontal black segment is drawn at $R=-0.0238$. 
}
\label{fig_rem}
\end{figure*}

The probability distribution for the parameter $R_L^{b}$ obtained by the bootstrap
procedure at $W=1240$~MeV ($E_\gamma=350$~MeV) is shown in the left panel of Fig.~\ref{fig_rem}.
As can be seen from the result of the fit, this distribution
can be considered approximately Gaussian, even though it
is given by the ratio of two non-Gaussian variables.

In the right panel of Fig.~\ref{fig_rem},
$R_L$ is plotted as a function of the center-of-mass energy $W$
for the upper part of the measured photon energy interval.
The quoted fit errors are the 68\% CL, and include the contribution of both
the statistical and angular-dependent systematic uncertainties of the experimental data. 

In the same figure, 
the $W$ dependence of the ratio: 
\begin{equation}\label{eq:rem-multi}
R_{W} =  \frac{\hbox{Im}[E_{1+}^{3/2}]}{\hbox{Im}
[M_{1+}^{3/2}]} \ ,
\end{equation}
as predicted by BOGA-2019, SAID-SM22 and DMT-2001,
is also shown. 
For all the different predictions, $R_W$ 
is strongly dependent on the $W$ value, while the 
experimental
$R_{L}$ values show quite small variations around the $M_\Delta$ value.

In order to give an estimate of $R_{EM}$ from the distribution of $R_L$,
a small center-of-mass energy interval $ W = 1232\pm 10$~MeV was then chosen,
and the weighted average of all $R$ values included in this interval was evaluated.
This interval is 
centered around the $M_\Delta$ value,
where $R_L$ can be considered constant and the approximated Eq.~(\ref{eq:approx})
holds.
The model systematic uncertainty associated with the approximations that led to Eq.~(\ref{eq:rem})
was evaluated, using the multipoles from both the previous  PW analyses and the 
DMT-2001 model, by
comparing the exact $R_{EM}$ value obtained with 
Eq.~(\ref{eq:1new}) with the average of the $R$ values calculated
with the approximated formula of Eq.~(\ref{eq:rem}) 
in the interval $W = 1232\pm 10$~MeV.  
The maximum relative deviation between these two estimates was found to be 4\%.

A 30\% variation in the width of the selected interval resulted in a $\sim 4\%$ change in both the
evaluated $R_{EM}$ value and its model systematic uncertainty.  
We then conservatively assume that the overall relative systematic uncertainty associated with both the model approximation and the interval choice is $\pm 4\%$ in rms units.

Finally, our new estimate of the parameter $R_{EM}$ is then:
\begin{equation}
  R_{EM}= [ -2.38 \pm 0.16 \hbox{(stat.+sys.)} \pm 0.10 \hbox{(model)} ] \% \ .
\end{equation}

\begin{figure*}%
\centering
\includegraphics[scale=0.3]{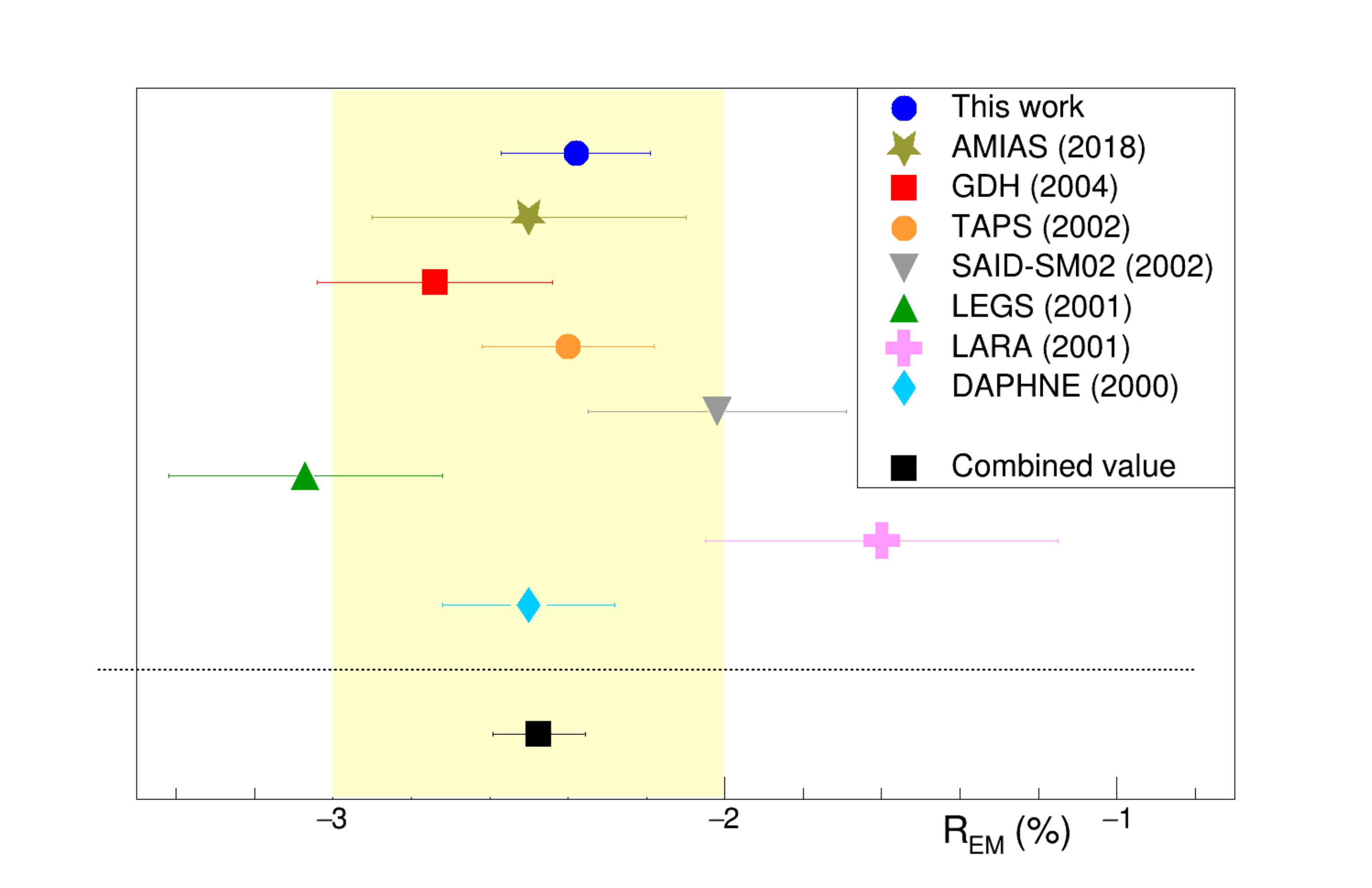}
\caption{The $R_{EM}$ value obtained in  this work (blue circle) is compared to the most recent evaluations from 
  Ref.~\cite{amias} (AMIAS - olive yellow star),    
  Ref.~\cite{ahr04} (GDH - red square),
  Ref.~\cite{beck3} (TAPS - orange circle),
  Ref.~\cite{SAID1} (SAID-SM02 - gray down triangle),
  Ref.~\cite{sand} (LEGS - green up triangle),
  Ref.~\cite{galler} (LARA - pink cross),
  Ref.~\cite{beck2} (DAPHNE - cyan diamond).
%
  %
  All quoted errors are the sum of both
  statistical, systematic, and model-dependent uncertainties.
  The vertical yellow band indicates the 100\% CL interval
  estimated by the PDG~\cite{PDG22}. 
  The black square below the dotted line is the newly calculated weighted average $R_{EM}^{AV}$,
  using some of the previous results (see text for details).}
\label{fig_remconf}
\end{figure*}

In Fig.~\ref{fig_remconf}, the newly evaluated $R_{EM}$ value is
compared to other recently published extractions given by the 
DAPHNE~\cite{beck2}, 
LARA~\cite{galler}, 
LEGS ~\cite{sand}, 
TAPS~\cite{beck3}, and
GDH~\cite{ahr04} collaborations,
by the SAID-SM02 PW analysis~\cite{SAID1}, and by the
AMIAS~\cite{amias} calculation. 
In this last approach, the $\gamma N \to N\pi$ reactions were described using 
the multipole expansions of the Chew Goldberger Low Nambu (CGLN) amplitudes~(\cite{cgln}).
A random set of multipoles coefficients was then generated by taking into account the
allowed physical limits and the required constraints. The predictions for the different observables given by each set of coefficients were finally compared to the selected $\gamma N \to N\pi$ experimental 
data set~(see Ref.\cite{amias}) and the final result and its uncertainty were evaluated 
from the obtained $\chi^2$ probability values.

The value of the systematic uncertainty associated to 
  the SAID-SM02 PW analysis ($\pm 3\%$) was taken from the  work of the
  BRAG Group~\cite{brag} that performed different multipole analyses using a commmon data base to gauge the model-dependence of such fits.

The present work provides an estimate with uncertainties that are not only more favorable, but also more rigorously calculated compared to previous work.

Within the quoted errors,
there is a good agreement between all the different evaluations. 
For this reason, it is possible to combine them to obtain
the most accurate extraction of $R_{EM}$ available, which should be used for any further reference
and for model comparison.

For this estimate,
we did not include data from the LEGS~\cite{sand}
collaboration because,
as the authors themselves stated in their original publication, a part of the data set from the 
$\gamma p \to N\pi$ reactions used for their evaluation is not consistent
with the data sets obtained at MAMI
(see also Refs.~\cite{beck1,beck2,beck3,ahr04}).
Similarly, we have also not included the $R_{EM}$ value from the LARA
collaboration~\cite{galler} since their proton Compton scattering data
give inconsistent results when compared to all other available data for this reaction (see Ref.~\cite{Mornacchi:2022}).

Our new evaluation was then calculated by combining the present value with those from Refs.~\cite{ahr04,beck2,beck3},
all based on independent data sets.
The weighted average of the selected values is:
\begin{equation}\label{eq:remval} 
  R_{EM}^{AV}= [ -2.47 \pm 0.12 \hbox{(stat.+sys.+mod.)} ] \% \  ,
\end{equation}
where the quoted error is the sum of the statistical, systematic, and model-dependent errors.


\section{Summary and conclusions}\label{sec:sum}
New precise data on the helicity-dependent differential cross section
of single $\pi^0$ photoproduction on the proton have been obtained. 
Compared to the existing data, this new measurement covers a  much larger energy
and polar angular interval with a significantly improved precision.
These new  data  
improve our understanding of the $p\pi^0$ photoproduction process in the $\Delta(1232)$ energy region, and, in particular, of the role of the low-lying multipoles, whose predicted contribution differs significantly between the available PW analyses and models.

From a Legendre moment analysis of the obtained angular distributions, the ratio
$ R_{EM}= (-2.38 \pm 0.16 \pm 0.10)\% $ at the $\Delta(1232)$ resonance mass value was obtained.
This is the most accurate estimate from one single experiment ever published until now.

By combining some of the available estimates, 
the most accurate experimental $R_{EM}$ value has been determined to be
$ R_{EM}^{AV}= (-2.47 \pm 0.12)\% $, where the quoted error is the sum of 
statistical, systematic, and model-dependent errors.
This value should be used for all further reference and model comparisons.

\section{Acknowledgments}

The authors wish to acknowledge the excellent support of the accelerator group of MAMI. 
This work was supported by SchweizerischerNationalfonds (200020-156983,132799, 121781,117601), DeutscheForschungsgemeinschaft (SFB443,SFB1044,SFB/TR16,FOR5327), the INFN-Italy, the European Community-Research Infrastructure Activity under FP7 programme (HadronPhysics, grantagreement No.227431), the U.K.Science and Technology Facilities Council grants ST/J000175/1, ST/G008604/1, ST/G008582/1,ST/J00006X/1, ST/V001035/1ST/V002570/1, ST/P004385/2, ST/T002077/1, and ST/L00478X/2, the Natural Sciences and Engineering Research Council (NSERC, FRN:SAPPJ-2015-00023), Canada. This material
is based upon work also supported by the U.S.Department of Energy, Office of Science, Office of Nuclear Physics Research Division, under Award Numbers DE-FG02-99ER41110, DE-FG02-88ER40415, DE-FG02-01-ER41194, DE-SC0014323andDE-SC0016583 and by the National Science Foundation, under Grant Nos.PHY-1039130 and IIA-1358175.
%

\appendix
\section{}
\begin{table*}
\caption{\label{tabref}
 Numerical values of the fitted Legendre coeffcients $(a_1)_0$, $(a_1)_1$, $(a_1)_2$  as a function of the total center-of-mass energy $W$.
  The quoted fit errors are the 68\% confidence level (CL) determined using the quantiles
of the bootstrapped parameter distributions, 
and include the contribution of both the statistical and systematic uncertainties of the experimental data. 
}
\begin{ruledtabular}
\begin{tabular}{c|c|c|c||c|c|c|c||c|c|c|c}
  $W$ & $(a_1)_0$ & $(a_1)_1$ & $(a_1)_2$ &   $W$ & $(a_1)_0$ & $(a_1)_1$ & $(a_1)_2$ &   $W$ & $(a_1)_0$ & $(a_1)_1$ & $(a_1)_2$ \\
  (MeV)  & ($\mu$b/sr) &  ($\mu$b/sr) &  ($\mu$b/sr) & (MeV)  & ($\mu$b/sr) &    ($\mu$b/sr) &  ($\mu$b/sr) & (MeV)  & ($\mu$b/sr) &  ($\mu$b/sr) &  ($\mu$b/sr) \\\hline
1079.5 & -1.0 $\pm$ 1.1 & 0.6 $\pm$ 1.9 & 0.8 $\pm$ 2.3 & 1159.0 & -4.4 $\pm$ 0.4 & -0.9 $\pm$ 0.5 & 10.8 $\pm$ 1.0 & 1226.1 & -15.1 $\pm$ 1.1 & 1.4 $\pm$ 0.5 & 26.0 $\pm$ 1.9 \\
1081.7 & 0.8 $\pm$ 0.8 & 1.9 $\pm$ 1.4 & 2.0 $\pm$ 1.7 & 1160.9 & -5.0 $\pm$ 0.5 & -0.2 $\pm$ 0.5 & 11.2 $\pm$ 1.0 & 1227.7 & -15.4 $\pm$ 1.2 & 1.9 $\pm$ 0.6 & 26.2 $\pm$ 2.0 \\
1083.9 & 0.7 $\pm$ 0.7 & 1.2 $\pm$ 1.2 & 1.0 $\pm$ 1.4 & 1162.8 & -5.7 $\pm$ 0.5 & -1.0 $\pm$ 0.5 & 11.5 $\pm$ 1.0 & 1229.2 & -14.9 $\pm$ 1.1 & 1.9 $\pm$ 0.5 & 25.7 $\pm$ 1.9 \\
1086.1 & 0.5 $\pm$ 0.6 & 0.2 $\pm$ 1.0 & 0.5 $\pm$ 1.3 & 1164.8 & -5.6 $\pm$ 0.5 & -0.3 $\pm$ 0.5 & 12.7 $\pm$ 1.1 & 1230.7 & -15.0 $\pm$ 1.1 & 1.3 $\pm$ 0.4 & 26.3 $\pm$ 2.0 \\
1088.3 & 0.5 $\pm$ 0.6 & -0.4 $\pm$ 0.9 & 1.6 $\pm$ 1.2 & 1166.7 & -6.3 $\pm$ 0.5 & -0.1 $\pm$ 0.5 & 13.9 $\pm$ 1.1 & 1232.3 & -14.3 $\pm$ 1.1 & 1.9 $\pm$ 0.5 & 25.1 $\pm$ 1.9 \\
1090.5 & -0.0 $\pm$ 0.5 & 0.1 $\pm$ 0.8 & 0.6 $\pm$ 1.0 & 1168.6 & -6.9 $\pm$ 0.6 & -0.5 $\pm$ 0.5 & 14.3 $\pm$ 1.2 & 1233.8 & -14.4 $\pm$ 1.1 & 2.3 $\pm$ 0.5 & 25.6 $\pm$ 1.9 \\
1092.6 & -0.4 $\pm$ 0.5 & -0.8 $\pm$ 0.8 & 1.7 $\pm$ 1.0 & 1170.5 & -7.5 $\pm$ 0.6 & 0.1 $\pm$ 0.5 & 15.3 $\pm$ 1.2 & 1235.3 & -13.6 $\pm$ 1.0 & 1.4 $\pm$ 0.5 & 23.9 $\pm$ 1.8 \\
1094.8 & 0.3 $\pm$ 0.6 & 0.5 $\pm$ 1.0 & -0.2 $\pm$ 1.3 & 1172.4 & -7.8 $\pm$ 0.6 & 0.4 $\pm$ 0.4 & 15.8 $\pm$ 1.2 & 1236.8 & -13.4 $\pm$ 1.0 & 2.3 $\pm$ 0.5 & 24.1 $\pm$ 1.8 \\
1097.0 & -0.2 $\pm$ 0.6 & -0.2 $\pm$ 0.9 & 0.6 $\pm$ 1.2 & 1174.3 & -8.1 $\pm$ 0.6 & 0.6 $\pm$ 0.5 & 15.3 $\pm$ 1.2 & 1238.2 & -13.7 $\pm$ 1.0 & 1.1 $\pm$ 0.5 & 23.6 $\pm$ 1.8 \\
1099.1 & -0.2 $\pm$ 0.9 & -1.5 $\pm$ 1.4 & 0.3 $\pm$ 1.7 & 1176.2 & -8.7 $\pm$ 0.7 & 0.4 $\pm$ 0.5 & 17.5 $\pm$ 1.4 & 1239.7 & -13.2 $\pm$ 1.0 & 2.4 $\pm$ 0.5 & 22.8 $\pm$ 1.7 \\
1101.3 & 0.2 $\pm$ 0.4 & 0.0 $\pm$ 0.6 & 1.3 $\pm$ 0.7 & 1178.0 & -9.2 $\pm$ 0.7 & 0.5 $\pm$ 0.5 & 18.1 $\pm$ 1.4 & 1241.1 & -12.7 $\pm$ 0.9 & 2.2 $\pm$ 0.5 & 23.2 $\pm$ 1.7 \\
1103.4 & 0.2 $\pm$ 0.5 & 0.9 $\pm$ 0.8 & 0.1 $\pm$ 1.0 & 1179.9 & -9.7 $\pm$ 0.7 & -0.1 $\pm$ 0.5 & 18.6 $\pm$ 1.4 & 1242.6 & -12.7 $\pm$ 1.0 & 1.5 $\pm$ 0.5 & 23.1 $\pm$ 1.8 \\
1105.6 & -0.2 $\pm$ 0.5 & -0.4 $\pm$ 0.9 & -0.5 $\pm$ 1.0 & 1181.7 & -10.2 $\pm$ 0.8 & 0.6 $\pm$ 0.5 & 19.6 $\pm$ 1.5 & 1244.0 & -12.2 $\pm$ 0.9 & 2.4 $\pm$ 0.6 & 21.8 $\pm$ 1.7 \\
1107.7 & 0.4 $\pm$ 0.8 & -0.1 $\pm$ 1.4 & 1.6 $\pm$ 1.7 & 1183.6 & -10.6 $\pm$ 0.8 & 0.2 $\pm$ 0.5 & 20.5 $\pm$ 1.6 & 1245.4 & -12.1 $\pm$ 0.9 & 1.8 $\pm$ 0.6 & 21.7 $\pm$ 1.7 \\
1109.9 & -0.5 $\pm$ 0.3 & -0.4 $\pm$ 0.6 & 1.7 $\pm$ 0.7 & 1185.4 & -11.1 $\pm$ 0.8 & 1.1 $\pm$ 0.4 & 21.1 $\pm$ 1.6 & 1246.8 & -12.2 $\pm$ 1.2 & 1.0 $\pm$ 0.7 & 21.8 $\pm$ 2.2 \\
1112.0 & -0.8 $\pm$ 0.3 & -0.5 $\pm$ 0.5 & 1.8 $\pm$ 0.7 & 1187.2 & -11.3 $\pm$ 0.9 & 0.9 $\pm$ 0.5 & 21.4 $\pm$ 1.6 & 1248.2 & -11.3 $\pm$ 0.9 & 2.0 $\pm$ 0.6 & 20.8 $\pm$ 1.6 \\
1114.1 & -1.2 $\pm$ 0.4 & -1.0 $\pm$ 0.6 & 1.5 $\pm$ 0.8 & 1189.0 & -12.2 $\pm$ 0.9 & 1.0 $\pm$ 0.5 & 22.8 $\pm$ 1.7 & 1249.6 & -10.9 $\pm$ 0.8 & 2.3 $\pm$ 0.6 & 19.8 $\pm$ 1.6 \\
1116.2 & -0.7 $\pm$ 0.3 & 0.0 $\pm$ 0.5 & 1.8 $\pm$ 0.7 & 1190.8 & -12.3 $\pm$ 0.9 & 1.0 $\pm$ 0.5 & 24.7 $\pm$ 1.8 & 1251.0 & -11.1 $\pm$ 0.9 & 2.2 $\pm$ 0.8 & 20.8 $\pm$ 1.7 \\
1118.3 & -0.6 $\pm$ 0.3 & 0.1 $\pm$ 0.5 & 2.6 $\pm$ 0.7 & 1192.6 & -13.0 $\pm$ 1.0 & 0.5 $\pm$ 0.5 & 24.4 $\pm$ 1.8 & 1252.3 & -10.5 $\pm$ 0.8 & 2.1 $\pm$ 0.5 & 18.4 $\pm$ 1.4 \\
1120.4 & -1.5 $\pm$ 0.3 & -1.0 $\pm$ 0.6 & 3.4 $\pm$ 0.7 & 1194.4 & -13.2 $\pm$ 1.0 & 0.9 $\pm$ 0.5 & 24.6 $\pm$ 1.8 & 1253.7 & -9.8 $\pm$ 0.8 & 1.4 $\pm$ 0.7 & 18.1 $\pm$ 1.5 \\
1122.5 & -1.3 $\pm$ 0.3 & -0.6 $\pm$ 0.5 & 2.6 $\pm$ 0.6 & 1196.2 & -13.9 $\pm$ 1.0 & 0.7 $\pm$ 0.5 & 25.5 $\pm$ 1.9 & 1255.0 & -10.2 $\pm$ 0.8 & 1.7 $\pm$ 0.7 & 18.2 $\pm$ 1.5 \\
1124.6 & -1.5 $\pm$ 0.3 & -0.5 $\pm$ 0.5 & 3.7 $\pm$ 0.7 & 1197.9 & -13.9 $\pm$ 1.0 & 0.3 $\pm$ 0.4 & 25.3 $\pm$ 1.9 & 1256.3 & -10.2 $\pm$ 0.9 & 1.4 $\pm$ 0.8 & 18.3 $\pm$ 1.6 \\
1126.7 & -1.8 $\pm$ 0.4 & 0.6 $\pm$ 0.7 & 2.6 $\pm$ 0.8 & 1199.7 & -13.9 $\pm$ 1.0 & 1.6 $\pm$ 0.5 & 26.6 $\pm$ 2.0 & 1257.6 & -9.8 $\pm$ 1.0 & 1.0 $\pm$ 1.1 & 18.4 $\pm$ 1.8 \\
1128.8 & -1.4 $\pm$ 0.3 & 0.3 $\pm$ 0.6 & 5.1 $\pm$ 0.8 & 1201.4 & -14.5 $\pm$ 1.1 & 1.2 $\pm$ 0.4 & 27.2 $\pm$ 2.0 & 1258.9 & -9.1 $\pm$ 0.8 & 2.0 $\pm$ 0.7 & 17.8 $\pm$ 1.5 \\
1130.8 & -1.8 $\pm$ 0.3 & -1.3 $\pm$ 0.5 & 3.9 $\pm$ 0.7 & 1203.1 & -14.8 $\pm$ 1.1 & 0.6 $\pm$ 0.5 & 26.7 $\pm$ 2.0 & 1260.2 & -8.9 $\pm$ 0.7 & 1.5 $\pm$ 0.7 & 17.9 $\pm$ 1.5 \\
1132.9 & -1.8 $\pm$ 0.3 & -0.5 $\pm$ 0.5 & 3.9 $\pm$ 0.7 & 1204.8 & -14.9 $\pm$ 1.1 & 1.5 $\pm$ 0.5 & 27.8 $\pm$ 2.1 & 1261.5 & -8.5 $\pm$ 0.7 & 2.3 $\pm$ 0.6 & 16.3 $\pm$ 1.4 \\
1134.9 & -2.4 $\pm$ 0.4 & -0.2 $\pm$ 0.6 & 5.3 $\pm$ 0.8 & 1206.5 & -15.0 $\pm$ 1.1 & 1.5 $\pm$ 0.5 & 26.6 $\pm$ 2.0 & 1262.8 & -8.7 $\pm$ 0.7 & 1.7 $\pm$ 0.7 & 15.0 $\pm$ 1.3 \\
1137.0 & -1.9 $\pm$ 0.4 & -0.5 $\pm$ 0.6 & 5.1 $\pm$ 0.8 & 1208.2 & -15.5 $\pm$ 1.1 & 1.8 $\pm$ 0.5 & 28.3 $\pm$ 2.1 & 1264.0 & -8.9 $\pm$ 0.7 & 1.7 $\pm$ 0.6 & 15.3 $\pm$ 1.3 \\
1139.0 & -2.6 $\pm$ 0.3 & 0.1 $\pm$ 0.5 & 5.5 $\pm$ 0.7 & 1209.9 & -15.6 $\pm$ 1.2 & 1.6 $\pm$ 0.7 & 27.8 $\pm$ 2.1 & 1265.2 & -8.4 $\pm$ 0.7 & 2.4 $\pm$ 0.7 & 16.0 $\pm$ 1.4 \\
1141.0 & -2.6 $\pm$ 0.5 & 0.7 $\pm$ 0.8 & 6.7 $\pm$ 1.1 & 1211.6 & -15.9 $\pm$ 1.2 & 1.1 $\pm$ 0.5 & 28.2 $\pm$ 2.1 & 1266.5 & -7.9 $\pm$ 0.7 & 3.3 $\pm$ 0.7 & 15.8 $\pm$ 1.4 \\
1143.1 & -2.4 $\pm$ 0.3 & -0.5 $\pm$ 0.5 & 6.4 $\pm$ 0.8 & 1213.2 & -16.0 $\pm$ 1.2 & 1.2 $\pm$ 0.4 & 28.5 $\pm$ 2.1 & 1267.7 & -7.9 $\pm$ 0.7 & 1.1 $\pm$ 0.7 & 13.4 $\pm$ 1.2 \\
1145.1 & -2.6 $\pm$ 0.3 & -0.4 $\pm$ 0.5 & 7.4 $\pm$ 0.8 & 1214.9 & -16.1 $\pm$ 1.2 & 1.2 $\pm$ 0.4 & 28.9 $\pm$ 2.1 & 1268.9 & -7.3 $\pm$ 0.8 & 1.0 $\pm$ 1.0 & 13.8 $\pm$ 1.6 \\
1147.1 & -3.0 $\pm$ 0.4 & -0.3 $\pm$ 0.6 & 8.7 $\pm$ 1.0 & 1216.5 & -16.1 $\pm$ 1.2 & 0.5 $\pm$ 0.5 & 28.4 $\pm$ 2.1 & 1270.1 & -7.3 $\pm$ 0.6 & 2.0 $\pm$ 0.6 & 13.6 $\pm$ 1.2 \\
1149.1 & -2.9 $\pm$ 0.3 & -0.5 $\pm$ 0.5 & 7.5 $\pm$ 0.8 & 1218.2 & -16.0 $\pm$ 1.2 & 1.5 $\pm$ 0.4 & 28.7 $\pm$ 2.1 & 1271.3 & -7.2 $\pm$ 0.6 & 2.4 $\pm$ 0.6 & 13.9 $\pm$ 1.2 \\
1151.1 & -3.8 $\pm$ 0.4 & 0.1 $\pm$ 0.5 & 9.1 $\pm$ 0.9 & 1219.8 & -16.1 $\pm$ 1.2 & 2.3 $\pm$ 0.5 & 28.0 $\pm$ 2.1 & 1272.5 & -7.6 $\pm$ 0.7 & 1.1 $\pm$ 0.7 & 14.0 $\pm$ 1.3 \\
1153.1 & -3.7 $\pm$ 0.4 & -0.2 $\pm$ 0.5 & 9.1 $\pm$ 0.9 & 1221.4 & -16.1 $\pm$ 1.2 & 2.0 $\pm$ 0.6 & 28.2 $\pm$ 2.1 & 1273.7 & -6.6 $\pm$ 0.5 & 1.1 $\pm$ 0.4 & 13.4 $\pm$ 1.1 \\
1155.0 & -4.1 $\pm$ 0.4 & -1.0 $\pm$ 0.5 & 9.9 $\pm$ 0.9 & 1223.0 & -16.1 $\pm$ 1.2 & 1.9 $\pm$ 0.6 & 27.3 $\pm$ 2.1 & 1275.0 & -6.3 $\pm$ 0.6 & 1.9 $\pm$ 0.7 & 12.1 $\pm$ 1.2 \\
1157.0 & -4.4 $\pm$ 0.4 & -0.8 $\pm$ 0.5 & 10.4 $\pm$ 0.9 & 1224.5 & -15.6 $\pm$ 1.2 & 1.9 $\pm$ 0.5 & 27.5 $\pm$ 2.1 & 1276.3 & -6.6 $\pm$ 0.6 & 1.6 $\pm$ 0.6 & 11.7 $\pm$ 1.1 \\
%
\end{tabular}
\end{ruledtabular}
\end{table*}

\clearpage

\end{document}